%% file: main.tex
\documentclass[sigconf, nonacm]{acmart}

\usepackage{pvldb}% VLDB/PVLDB formatting overrides, see pvldb.sty for details.
\usepackage{enumitem}
\usepackage{stmaryrd}
\usepackage{tikz}
\usetikzlibrary{arrows.meta,positioning,calc}
\usepackage{listings}
\definecolor{gendbblue}{HTML}{536E9A}
\definecolor{gendbpaleblue}{HTML}{E7EEF9}
\definecolor{gendbgreen}{HTML}{4F7D5B}
\definecolor{gendbpalegreen}{HTML}{E0F0E5}
\definecolor{gendbgray}{HTML}{4C5C68}
\lstdefinelanguage{Verus}{
  morekeywords={fn,let,mut,while,if,return,requires,ensures,invariant,decreases,proof,as},
  sensitive=true,morecomment=[l]{//}
}
\lstdefinestyle{gendbcode}{
  language=Verus,basicstyle=\ttfamily\small,
  keywordstyle=\bfseries\color{gendbgray},commentstyle=\color{gendbgray},
  columns=fullflexible,keepspaces=true,showstringspaces=false,
  breaklines=true,aboveskip=3pt,belowskip=0pt,
  frame=single,rulecolor=\color{gendbgray!55},framesep=4pt
}

\renewcommand\vldbdoi{XX.XX/XXX.XX}
\renewcommand\vldbpages{XXX-XXX}
\renewcommand\vldbavailabilityurl{https://github.com/GenLang-V/GenDB-V}

\begin{document}
\title{GenDB-V: AI-Generated and Formally Verified Database Systems}

\author{Jiale Lao}
\affiliation{%
  \institution{Cornell University}
  \city{Ithaca}
  \state{New York}
  \country{USA}
}
\email{jiale@cs.cornell.edu}

\author{Immanuel Trummer}
\affiliation{%
  \institution{Cornell University}
  \city{Ithaca}
  \state{New York}
  \country{USA}
}
\email{itrummer@cornell.edu}

\begin{abstract}
\input{sections/00-abstract}
\end{abstract}

\maketitle

\vldbtopmatter

\input{sections/01-introduction}
\input{sections/02-overview}
\input{sections/03-sqlspec}
\input{sections/04-obligations}
\input{sections/05-synthesis}
\input{sections/06-evaluation}
\input{sections/07-related}
\input{sections/08-conclusion}

\balance

\bibliographystyle{ACM-Reference-Format}
\bibliography{gendb-v}

\end{document}

%% file: sections/00-abstract.tex
% Final template results: RESULTS_INDEX_2026-10-29.md / FT/summary.json, per-template arithmetic means over visible bindings, then geometric-mean speedups: 51.56x / 21.96x.
% Q08/Q14 Umbra numeric-scale differences qualify comparisons, not correctness; disclose in Evaluation.
% Final TPC-H author cost $1,867.93 includes storage and all optimization rounds; framework work excluded.
% Results (Section 6, checked 2026-11-30): robust plain-array TPC-H code 5.69x / 2.22x on the given database (RB1/given.json); lowest total time of all systems, Bespoke OLAP included, on the attack databases (213 s) and in each group of held-out databases (RB1/attack.json, RB1/heldout.json); joint 22.0x-61.3x; final single engine 1.09x-5.02x in total workload time at 1/32/64 threads on TPC-H and SEC-EDGAR.

For over 50 years, the database community has developed a wide range of successful engines to answer two fundamental questions: how can database systems run faster, and how to ensure their correctness? Experience with these engines shows that performance depends on hardware and workloads. As both change over time, we repeatedly redesign and rebuild database systems, a time-consuming and expensive process. Correctness remains another long-standing challenge: recent studies continue to uncover wrong-answer bugs even in mature database engines, despite years of development and testing.
To address these, we present GenDB-V, a system that enables AI agents to synthesize database engines specialized to hardware and workloads for performance, and formally verified for correctness. It supports specialized implementations of SQL templates and general-purpose engines.
% that accept any query within a supported SQL subset. 
First, we develop SQLSpec to formalize SQL semantics and deterministically construct fixed, implementation-agnostic synthesis targets for storage preservation and query correctness. Second, we enable joint synthesis of fast code and formal proofs through storage--execution co-optimization, performance-first exploration, three-layer modular proofs that bridge low-level execution to SQLSpec semantics, and robust code synthesis that generalizes verified code beyond the given database state. Third, we verify the synthesized code and proofs against the fixed synthesis targets in an isolated pipeline, proving SQL correctness and program safety.
We evaluate GenDB-V on TPC-H and SEC-EDGAR, a benchmark designed to reduce LLM training-data leakage. On TPC-H SF100, the robust verified code is 5.69$\times$ faster than DuckDB and 2.22$\times$ faster than Umbra, and it remains the fastest system on attack databases built by the agents and on held-out databases built independently. Joint storage and query optimization raises the speedups to 22.0--61.3$\times$ on both benchmarks, and a verified general-purpose engine outperforms both systems by 1.09--5.02$\times$. Our experiments expose incorrect answers in both a human-built engine and a testing-based LLM synthesis method.

%% file: sections/01-introduction.tex
\section{Introduction}
\label{sec:intro}

For more than five decades, the database community has developed successful engines for workloads ranging from short transactions~\cite{astrahan1976systemr,stonebraker2007era} to analytical queries~\cite{boncz2005x100,neumann2011compiling,neumann2020umbra}, and more~\cite{abadi2003aurora,stonebraker2011scidb}. Across these workloads, two fundamental questions have remained central: how can database systems run faster, and how can we ensure their correctness? Despite decades of progress, both remain long-standing challenges that continue to drive database research.

Database performance depends on hardware and workloads, and changes in both have repeatedly led to costly redesigns and reimplementations. In 2007, H-Store redesigned transaction processing for main memory and motivated a call for ``a complete rewrite''~\cite{stonebraker2007era}. This call has returned many times: Snowflake for the cloud~\cite{dageville2016snowflake}, DuckDB for embedded analytics~\cite{raasveldt2019duckdb}, Umbra for fast SSDs~\cite{neumann2020umbra}, and Sirius for GPUs~\cite{yogatama2025sirius}. Each new design addresses a new setting, but hardware and workloads continue to evolve.
% What will demand the next redesign, and how many more times must we rebuild?

Ensuring correctness remains difficult: testing continues to expose bugs, while formal verification requires substantial human effort. EET found 35 logic bugs across five database systems, including a PostgreSQL hash-join bug that had survived for 20 years~\cite{jiang2024eet}. More recently, SQLancer++ reported 196 bugs across 18 systems, including 140 logic bugs~\cite{zhong2026scaling}. These findings illustrate why extensive testing cannot establish correctness for every valid input. Formal verification can establish that code satisfies a precise specification for every input covered by the proof. For example, CompCert proves that compilation preserves program behavior~\cite{leroy2009compcert}. In a compiler-testing study, Csmith reported 79 bugs in GCC and 202 in LLVM, but found no wrong-code bugs in CompCert's verified passes despite approximately six CPU-years of testing~\cite{yang2011csmith}. However, developing such proofs requires substantial effort: verifying seL4's 8,700-line C kernel required approximately 20 person-years~\cite{klein2009sel4}.

Recent advances in large language models (LLMs) have enabled agents to optimize software and generate machine-checked proofs~\cite{novikov2025alphaevolve,yang2026verusage}. AlphaEvolve~\cite{novikov2025alphaevolve} improved production code at Google beyond expert-designed implementations, while GSO~\cite{shetty2025gso, gso2026leaderboard} demonstrates agents' growing ability to match expert optimizations across real software repositories. Progress in proof generation spans both software and mathematics: VeruSAGE~\cite{yang2026verusage} constructs correctness proofs for systems software, AlphaProof~\cite{hubert2026alphaproof} produces Lean proofs for Olympiad problems, and Gauss~\cite{mathincgauss} helped formalize the strong prime number theorem in Lean.
% ~\cite{yang2026verusage,hubert2026alphaproof,mathincgauss}.

Building on these advances, we present GenDB-V to address the performance and correctness challenges of database systems by having AI agents synthesize engines that are specialized to hardware and workloads for performance and formally verified for correctness. Our key insight is to deterministically construct fixed formal synthesis targets, let agents freely explore optimizations and proofs together, and independently verify that their implementations satisfy those targets. The targets define the required behavior without prescribing how it must be implemented. Agents choose storage layouts and execution strategies, then develop intermediate contracts and proofs connecting their code to these targets. An independent verifier checks both that the code satisfies the intermediate contracts and that those contracts establish the fixed requirements.

First, we develop SQLSpec to formalize SQL semantics and deterministically construct fixed, implementation-agnostic synthesis targets for storage preservation and query correctness. SQLSpec follows PostgreSQL semantics and accounts for nulls, duplicate rows, runtime SQL errors, and the multiple valid results that SQL permits. GenDB-V generates the targets outside the agent's control, so agents cannot hallucinate or weaken them. The targets specify only input assumptions and permitted outcomes: storage builders must preserve all input data, and the synthesized code must return a permitted result on every valid database, for all allowed parameter bindings of a SQL template or for all supported queries of a general-purpose engine.

Second, we enable agents to jointly synthesize fast implementations and formal proofs. Agents co-design storage layouts and execution algorithms, and a representation predicate exposes the physical layout to the query proofs. Because proofs are expensive, agents first prototype and profile candidate designs without verification and develop proofs only for the fastest design. Agents decompose each proof into three modular layers: storage reader contracts, physical loop invariants, and bridge lemmas that connect physical execution to SQLSpec semantics. Finally, code that is optimized for one given database state can become slow when the data changes. In robust code synthesis, agents therefore construct valid databases and parameter bindings that expose weaknesses of the code, and revise the code and the proofs to remove them.

Third, we verify the synthesized code and proofs against the fixed synthesis targets in an isolated pipeline built on Verus~\cite{lattuada2023verus,verusguide}, proving SQL correctness, data preservation, memory and arithmetic safety, race-free parallel execution, and termination. All specification files and the top-level entry-point function are read-only for agents. Once all proofs check, the verified Rust code is compiled into a shared library for query execution.

We evaluate GenDB-V on TPC-H at SF100 and SEC-EDGAR at SF3, a benchmark designed to reduce LLM training-data leakage~\cite{lao2026gendb}. GenDB-V produces verified implementations for all 28 templates and a verified general-purpose engine. Our main setting synthesizes robust query code over fixed plain column arrays. On the given TPC-H database, this code achieves geometric-mean speedups of 5.69$\times$ over DuckDB and 2.22$\times$ over Umbra. We also measure it on attack databases that the agents built to expose weaknesses of their own code, and on 29 held-out databases that we created independently of the synthesis. On both, the robust code remains the fastest system in total time, whereas the code before robust code synthesis is very slow in 79\% and 24\% of the runs.

When agents also optimize the storage for one given database, under a policy that requires lossless base data and prohibits precomputed results, the speedups grow to 22.0--61.3$\times$ over DuckDB and Umbra across the two benchmarks. One verified general-purpose engine serves both benchmarks and accepts any well-typed query within SQLSpec's supported fragment. It evolves from a reference interpreter that completes only two TPC-H queries at SF1 within the timeout into an engine that outperforms DuckDB and Umbra by 1.09--5.02$\times$ in total workload time. GenDB-V also achieves lower execution times than Bespoke OLAP's testing-based LLM synthesis~\cite{wehrstein2026bespoke}. Our experiments expose incorrect answers in both human-built and testing-based systems: a wrong query result in Umbra 26.08\footnote{We evaluate this version because the \texttt{latest} tag of Umbra's Docker image points to it at the time of writing. The wrong result can be reproduced with the notebook at \url{https://colab.research.google.com/drive/1sGJOLsFNXI859cYUT_wS5eb-oYDqpPp7}.}, which later versions correct, and incorrect answers in Bespoke OLAP's implementations for 16 of 28 templates.

% Umbra version: on 2026-11-26 the `latest` tag of the image umbradb/umbra still pointed to 26.08 (GenDB-codex .lane-tmp/general-engine/UR1/HANDOFF.md).
% Costs (Section 6.7): plain-array TPC-H $1,285 + robust code synthesis $1,934 = $3,219; joint $1,868 + $442 = $2,310; engine $2,833. Very slow runs of the initial code: 595 of 750 (79%) on the attack databases, 1,215 of 5,060 (24%) on the held-out databases.
Synthesis costs \$3,219 for the robust TPC-H implementations, \$2,310 for joint optimization across both benchmarks, and \$2,833 for the general engine. Compared with the substantial human effort required by prior verification projects, these results demonstrate a promising path toward automating the development of fast, formally verified database systems. Future work should validate and expand SQLSpec, explore systematic and user-customizable robust code synthesis, and reduce synthesis costs.

% In building GenDB-V, we make the following contributions:

% The original scientific contributions are the following:
% \begin{itemize}[leftmargin=*,labelindent=0pt]
%     \item \textbf{SQLSpec and fixed formal synthesis targets.} We develop SQLSpec (defining mathematical SQL semantics and a verified reference interpreter) and deterministically construct fixed storage and execution targets for SQL templates and general-purpose query engines outside the agent's control (Section~\ref{sec:sqlspec}).
%     \item \textbf{Joint code and proof synthesis with independent verification.} We enable agents to synthesize fast, verified implementations through storage--execution co-optimization, performance-first exploration, three-layer modular proofs, and robust code synthesis, checked by an independent Verus verification pipeline in an isolated environment (Sections~\ref{sec:joint-synthesis}--\ref{sec:verification}).
%     \item \textbf{Experimental evaluation.} We evaluate verified template implementations on TPC-H and SEC-EDGAR and a verified general-purpose engine on TPC-H. Template implementations achieve geometric-mean speedups of 22.0--61.3$\times$ over DuckDB and Umbra, while the general engine achieves total-workload speedups of 1.27--3.4$\times$. We also analyze synthesis time and monetary cost, code and proof size, the effects of storage optimization, correctness beyond synthesis tests, and case studies (Section~\ref{sec:evaluation}).
% \end{itemize}

%% file: sections/02-overview.tex
\section{GenDB-V Overview}
\label{sec:overview}

\begin{figure}[t]
  \centering
  \includegraphics[width=\columnwidth]{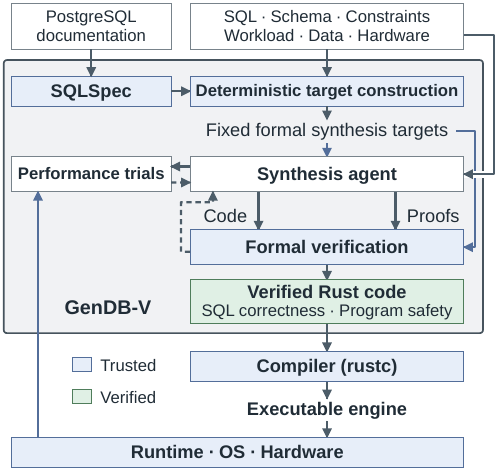}
  \caption{GenDB-V overview.}
  \label{fig:overview}
\end{figure}

Synthesizing a fast and formally verified database engine presents a core design challenge: to achieve high performance, agents need the freedom to invent specialized storage layouts and execution algorithms, yet an untrusted agent should not define or weaken the correctness requirement, and the correctness requirement should not restrict the agent's implementation choices. GenDB-V addresses this challenge by separating the definition and verification of correctness from the implementation search, and by specifying fixed formal synthesis targets purely in terms of required SQL behavior rather than physical implementation details. Before synthesis begins, GenDB-V deterministically constructs the fixed formal synthesis targets from the SQL workload and schema. During synthesis, agents freely explore physical implementations, intermediate contracts, and proofs together. Before admission, an independent verifier checks that the generated code and proofs establish the fixed formal synthesis targets. Figure~\ref{fig:overview} shows the system architecture.

% Blue marks trusted components and definitions; green marks verified code. Dashed arrows return verification and performance feedback.

First, deterministic target construction (Section~\ref{sec:sqlspec}) translates the input SQL, schema, and integrity constraints into fixed formal synthesis targets. It uses SQLSpec, our formal definition of SQL semantics in Verus. A target covers either one SQL template across all allowed parameter bindings, or a general-purpose engine across all queries within the supported SQL subset. When agents optimize storage, the targets also require storage builders and readers to preserve the logical data.

Second, joint code and proof synthesis (Section~\ref{sec:joint-synthesis}) lets agents explore implementations and proofs together. Guided by the workload, data, and hardware, agents co-design storage layouts and execution algorithms, or optimize query execution over fixed storage. They first prototype and profile candidate implementations without proofs, and then prove correctness for the fastest design in three modular layers that together establish the formal synthesis target. In robust code synthesis, agents construct valid databases and parameter bindings that expose weaknesses of an implementation, and revise the code and the proofs to make the code robust.

Third, independent verification (Section~\ref{sec:verification}) checks the generated code and proofs against the fixed formal synthesis targets without trusting the agent. If verification fails, GenDB-V returns the verifier's feedback to the agent. Once verification succeeds, the verified Rust code guarantees SQL correctness and program safety. The compiler then erases the proof code and compiles the verified Rust code into an executable engine, so query execution incurs no proof-checking overhead. As shown in Figure~\ref{fig:overview}, SQLSpec, deterministic target construction, the verifier, the compiler, and the runtime environment form the trusted foundation, whereas no agent-generated code or proof is trusted.

%% file: sections/03-sqlspec.tex
\section{Formal Synthesis Targets}
\label{sec:sqlspec}

GenDB-V formalizes SQL semantics in a reusable library, SQLSpec (Section~\ref{sec:sqlspec-semantics}), and deterministically constructs fixed formal synthesis targets from it before agent synthesis begins (Section~\ref{sec:target-construction}). Each target specifies only the input assumptions and permitted outcomes on every valid database, and leaves physical representations, execution algorithms, and proof strategies to the agent.
% Table~\ref{tab:notation} summarizes the main notation.

% \begin{table}[t]
% \centering\small
% \caption{Main notation for formal synthesis targets.}
% \label{tab:notation}
% \setlength{\tabcolsep}{3pt}
% \begin{tabular}{@{}l p{0.69\columnwidth}@{}}
% \toprule
% Symbol & Meaning \\
% \midrule
% $\Sigma$ & Schema: column types and integrity constraints \\
% $D,\ D[t]$ & Logical database and relation at table index $t$ \\
% $\mathsf{Valid}_\Sigma(D)$ & $D$ satisfies the types and constraints of $\Sigma$ \\
% $q,\ Q$ & Query body and top-level query with ordering, limit, and offset \\
% $\mathsf{eval}(D,\rho,q)$ & Evaluation of $q$ with outer-row context $\rho$ \\
% $\llbracket Q\rrbracket(D)$ & Set of SQLSpec outcomes permitted for $Q$ on $D$ \\
% $\mathsf{Rows}(R')$ & Successful outcome containing row sequence $R'$ \\
% $\mathsf{SqlError}(e)$ & Outcome reporting SQL error $e$ \\
% $T,\ p,\ \mathcal P$ & SQL template, parameter binding, and allowed binding domain \\
% $Q_T(p)$ & SQLSpec query for template $T$ with binding $p$ \\
% $\mathcal Q_\Sigma$ & Well-typed queries over $\Sigma$ in the supported subset \\
% $B,\ S=B(D)$ & Storage builder and its physical storage output \\
% $\mathsf{Rep}(S,D)$ & Representation predicate relating $S$ to $D$ \\
% $E_T(S,p)$ & Template executor on storage $S$ and binding $p$ \\
% $E(S,Q)$ & General-purpose engine on storage $S$ and query $Q$ \\
% $\mathsf{Allowed}(D,Q,o)$ & Outcome $o$ is permitted by SQLSpec or is a justified capacity refusal \\
% \bottomrule
% \end{tabular}
% \end{table}

\subsection{SQLSpec: Formal SQL Semantics}
\label{sec:sqlspec-semantics}

Formalizing SQL for database engine verification requires capturing real SQL behaviors that idealized relational algebra omits, including nulls, duplicate rows, runtime errors, and nondeterministic row ordering. We develop SQLSpec, a formal SQL library in Verus based on PostgreSQL documentation~\cite{postgresql18docs}. SQLSpec covers the SQL features needed by TPC-H and SEC-EDGAR (Table~\ref{tab:sqlspec-layers}). It can be extended with further features, and it rejects currently unsupported ones such as window functions. It consists of a logical data model and query syntax, mathematical evaluation functions that define the permitted outcomes of a query, and a verified executable interpreter.

\begin{table}[t]
\centering\small
\caption{Supported SQL subset and semantics in SQLSpec.}
\label{tab:sqlspec-layers}
\setlength{\tabcolsep}{3pt}
\begin{tabular*}{\columnwidth}{@{}l@{\extracolsep{\fill}}l@{}}
\toprule
Category & Supported features and semantics \\
\midrule
Schema & Typed values, \textsf{NULL}, primary/unique/foreign keys, \texttt{CHECK} \\
Expressions & Literals, columns, arithmetic, casts, \texttt{CASE}, scalar subqueries \\
Predicates & Comparisons, 3-valued logic, \texttt{EXISTS}/\texttt{IN}/\texttt{ANY}/\texttt{ALL} subqueries \\
Operators & \texttt{Table}, \texttt{Filter}, \texttt{Project}, \texttt{Product}, \texttt{LeftJoin}, \texttt{Distinct} \\
Aggregates & \texttt{COUNT}, \texttt{SUM}, \texttt{AVG}, \texttt{MIN}, \texttt{MAX}, \texttt{DISTINCT}, \texttt{GROUP BY} \\
Ordering & \texttt{ORDER BY}, \textsf{NULL} order, \texttt{OFFSET}, \texttt{LIMIT}, permitted tie orders \\
\bottomrule
\end{tabular*}
\end{table}

\noindent\textbf{Data model and query syntax.}
To keep the specification independent of physical storage layouts, SQLSpec defines a logical data model and represents queries as operator trees (Table~\ref{tab:sqlspec-layers}). In the data model, values are mathematical integers, exact decimals, dates, strings, booleans, or \textsf{NULL}. A row $r$ is a sequence of values, a relation $R$ is a sequence of rows that preserves duplicates, and a database $D$ contains one relation $D[t]$ for each table index $t$ (starting from 0) in the schema $\Sigma$. Note that the sequence representation of relations is a modeling choice and does not change SQL's bag semantics. To enforce schema rules and integrity constraints, SQLSpec defines boolean specification functions over $D$ that check whether every value fits its declared column type and nullability, and whether the relations satisfy the declared primary/unique/foreign keys, and \texttt{CHECK} constraints (deterministic conditions that each row must satisfy). These functions serve as preconditions in the formal synthesis targets, restricting the proof to valid databases. Over this data model, SQLSpec represents a SQL query by composing syntax constructors into an operator tree. Each constructor specifies a relational operator, expression, predicate, or aggregate together with its arguments and child queries. For example, \texttt{Query::Filter(p, q)} applies predicate $p$ to each row produced by child operator tree $q$, retaining rows for which $p$ evaluates to \textsf{True}.

% Queue the two-column figure early so it can appear beside Section 3.2.
\input{figures/sqlspec-example-figure}

\noindent\textbf{Mathematical evaluation and permitted outcomes.}
To give operator trees a mathematical meaning, SQLSpec defines a pure Verus specification function for each syntax constructor and evaluates a query tree recursively by composing these functions. Specification functions define mathematical behavior for proof checking and are erased during compilation, so they add no runtime overhead. To handle correlated subqueries, SQLSpec defines the outer-row context $\rho$ as a flat sequence of values from the current row at each enclosing query level, where a row may combine multiple tables. The frontend resolves names and aliases to positional column references before evaluation. An evaluation function $\mathsf{eval}(D,\rho,q)$ evaluates an operator tree $q$ on database $D$ and outer-row context $\rho$, recursively evaluating child queries and applying each constructor's specification function to return either a relation or a specified SQL error. For example, evaluating $\texttt{Query::Filter}(p,q)$ first calls $\mathsf{eval}(D,\rho,q)$ to evaluate child query $q$, and applies the filter specification function to evaluate predicate $p$ on each row under three-valued logic (\textsf{True}, \textsf{False}, \textsf{Unknown}): if either $q$ or $p$ raises a runtime error (such as division by zero), evaluation returns that SQL error; otherwise, it keeps only rows where $p$ evaluates to \textsf{True}. The remaining constructors formalize standard SQL semantics, so we omit their detailed definitions for brevity.

A SQL query can permit multiple valid results due to unspecified row order, ordering ties, and limits. Requiring a single canonical result could reject correct results produced by valid execution plans. For a top-level query $Q=(q,\vec o,\ell,d)$ with body $q$, ordering keys $\vec o$, optional limit $\ell$, and offset $d$, SQLSpec defines the set of permitted outcomes $\llbracket Q\rrbracket(D)$. We write $\mathsf{Rows}(R')$ for a successful outcome containing row sequence $R'$, and $\mathsf{SqlError}(e)$ for an outcome reporting SQL error $e$. If $\mathsf{eval}(D,[],q)$ returns error $e$, the only permitted outcome is $\mathsf{SqlError}(e)$. If evaluation returns relation $R$, SQLSpec defines permitted outcomes in two steps. First, SQLSpec considers permitted orderings of all rows in $R$, preserving duplicates. Without \texttt{ORDER BY}, every ordering is permitted. With \texttt{ORDER BY}, SQLSpec permits every ordering that follows the sort keys, including every permutation of rows with equal sort keys. Second, SQLSpec applies \texttt{OFFSET} followed by \texttt{LIMIT} to each permitted ordering. Each resulting sequence $R'$ gives a permitted outcome $\mathsf{Rows}(R')$. Query correctness requires the returned outcome to belong to $\llbracket Q\rrbracket(D)$.

\noindent\textbf{Verified executable interpreter.}
In Verus, specification functions exist only for verification: they operate over unbounded mathematical integers and sequences and are erased by the compiler, so they cannot be compiled into a runnable program. SQLSpec therefore pairs its specification with an executable reference interpreter, written in Rust over fixed-width integers and byte vectors. The interpreter is proved in Verus to satisfy $\llbracket Q\rrbracket(D)$ for every valid database $D$ and query $Q$ (Sections~\ref{sec:target-construction} and~\ref{sec:modular-proofs}). GenDB-V uses it as the initial engine that agents optimize into a high-performance general-purpose engine.

\subsection{Deterministic Target Construction}
\label{sec:target-construction}

GenDB-V supports two synthesis modes: \emph{template specialization}, which synthesizes a query executor for one parameterized SQL template across all allowed parameter bindings, and \emph{engine synthesis}, which synthesizes a general-purpose query engine across all queries within the supported SQL subset. If an agent translated SQL into the formal theorem, hallucination or reward hacking could misinterpret SQL semantics, drop constraints, or restrict the input domain. GenDB-V therefore translates SQL templates, database schemas, and the supported SQL subset into fixed formal synthesis targets deterministically, outside the agent's control.

\noindent\textbf{From SQL text to SQLSpec terms.}
First, the frontend parses a SQL query with SQLGlot~\cite{sqlglot} and resolves table and column names, expression types, and type conversions against the schema under PostgreSQL rules to produce a typed AST (Figure~\ref{fig:sqlspec-example}, steps 1--2). Next, a deterministic translator converts this typed AST into a SQLSpec query tree (step~3). Because this translator is part of the trusted foundation, SQLSpec is designed so that its syntax constructors directly mirror the nodes of a typed SQL AST. The translator can therefore map each relational operator, expression, predicate, and aggregate in the AST one-to-one to its corresponding SQLSpec constructor, keeping the translator small, auditable, and easy to extend. In template specialization, a SQL template $T$ is fixed at synthesis time and may contain parameters (\texttt{\$1}, \texttt{\$2}, \dots), each with a resolved SQL type and an allowed value range. For any parameter binding $p$ within these ranges, the translator replaces each parameter \texttt{\$i} with its value $p_i$ to form a parameterized SQLSpec query tree $Q_T(p)$. In engine synthesis, the query is not fixed ahead of time: instead, the translator converts each incoming SQL query at runtime into a SQLSpec query tree $Q$, and the synthesized engine must handle any valid query tree $Q$ within the supported SQL subset.

\begin{example}
\label{ex:running-sqlspec}
Figure~\ref{fig:sqlspec-example} illustrates this pipeline on our running example over table \texttt{readings} (with column \texttt{value BIGINT NOT NULL} and constraint \texttt{CHECK(value >= 0)}) for template $T$ (step~1):
\begin{center}\small\ttfamily
SELECT COUNT(*) AS n FROM readings WHERE value > \$1;
\end{center}
First, the frontend resolves table \texttt{readings} and column \texttt{value}, infers that \texttt{value} (\texttt{BIGINT NOT NULL}) and parameter \texttt{\$1} (\texttt{BIGINT}) both have 64-bit integer type, and builds a typed AST with bottom-to-top operator nodes \textsf{Table: readings}, \textsf{Filter: value $>$ \$1}, and \textsf{Aggregate: COUNT(*)} (step~2). Next, for any parameter binding $p=(k)$ where $k$ lies in the range $[-2^{63},2^{63}-1]$, the translator maps each AST node to its corresponding SQLSpec constructor to form $Q_T(k)$ (step~3): \texttt{Query::\allowbreak Table(0)} for table~0, \texttt{Query::\allowbreak Filter} with predicate $\texttt{Col}(0)>\texttt{Lit}(k)$, and \texttt{Query::\allowbreak Aggregate} over \texttt{[Agg::\allowbreak CountStar]}. Finally, SQLSpec's mathematical semantics evaluates $Q_T(k)$ bottom-to-top on database $D$ (step~4). Let $R=\mathsf{table}_0(D)=D[0]$ be the rows of table~0 and $x(r)$ be the integer at column~0 of row $r$. The filter keeps all qualifying rows, preserving duplicates:
\begin{equation}
F_k(D)=[\,r\in R\mid x(r)>k\,].
\label{eq:example-filter}
\end{equation}
The aggregate then counts the qualifying rows. When $|F_k(D)|\le 2^{63}-1$, the unique permitted outcome in $\llbracket Q_T(k)\rrbracket(D)$ is the single-row relation $\mathsf{Rows}([[|F_k(D)|]])$; when $|F_k(D)|>2^{63}-1$, the unique permitted outcome is $\mathsf{SqlError}(22003)$. On example input $D[0]=[[5],\allowbreak[5],\allowbreak[9],\allowbreak[0]]$ with $k=4$, the filter keeps $F_4(D)=[[5],\allowbreak[5],\allowbreak[9]]$ and the required output is $[[3]]$.
\end{example}

\noindent\textbf{Storage targets: builder correctness and losslessness.}
The agent defines a storage builder $B$ and a representation predicate $\mathsf{Rep}(S,D)$ that states whether physical storage $S$ represents logical database $D$. $\mathsf{Rep}$ is a Verus specification function whose definition depends on the chosen layout, such as compressed columns, reordered rows, or zone maps. GenDB-V leaves its definition to the agent, because a fixed definition would restrict storage designs, and requires the agent to prove two fixed targets.

First, \emph{builder correctness} requires building storage from a valid database establishes $\mathsf{Rep}$. With $\mathsf{Valid}_\Sigma(D)$ denoting that $D$ satisfies the schema types and integrity constraints of $\Sigma$, the target is:
\begin{equation}
\forall D:\quad \mathsf{Valid}_\Sigma(D) \Longrightarrow \mathsf{Rep}(B(D),D).
\label{eq:storage-builder}
\end{equation}
Second, because an untrusted agent defines $\mathsf{Rep}$, builder correctness alone does not guarantee that the storage preserves the data. \emph{Losslessness} therefore requires that the same physical storage cannot represent two different logical databases:
\begin{equation}
\forall S,D_1,D_2:\quad
\mathsf{Rep}(S,D_1)\land\mathsf{Rep}(S,D_2)
\Longrightarrow D_1=D_2.
\label{eq:storage-lossless}
\end{equation}
Together, the two targets imply that $B$ is injective on valid databases. By the left-inverse theorem~\cite{plp-inverse-functions,ernst-left-inverse,mathlib-left-inverse}, a recovery function $G$ with $G(B(D))=D$ for every valid $D$ therefore exists. Here, $D_1=D_2$ requires the same rows in the same order in each logical table, so storage may reorder rows but must keep enough information to recover the logical order. Query execution proofs use the same $\mathsf{Rep}(S,D)$ to show that execution over $S$ returns an outcome that SQLSpec permits on $D$.

\noindent\textbf{Query and engine execution targets.}
Both synthesis modes use the same outcome condition to connect physical query execution over $S$ to SQLSpec semantics over $D$. Because executable code uses fixed-width numbers (such as 64-bit integers and 128-bit decimals) and byte vectors for strings, we define $\mathsf{Allowed}(D,Q,o)$ for a query tree $Q$ and execution outcome $o$ to hold if and only if $o\in\llbracket Q\rrbracket(D)$, or $o=\mathsf{Capacity}$ when evaluating $Q$ on $D$ produces an intermediate number that exceeds its type bounds.

In \emph{template specialization}, the SQL template $T$ is fixed at synthesis time and the synthesized executor $E_T$ receives a tuple of parameter values $p$ at runtime. Let $\mathcal{P}$ denote the declared domain of allowed parameter bindings for $T$. The fixed formal synthesis target requires $E_T$ to return a permitted outcome for every valid database $D$, every matching physical storage state $S$, and every allowed parameter binding $p\in\mathcal{P}$:
\begin{equation}
\begin{split}
\forall D,S,\ \forall p\in\mathcal{P}:\quad
&\mathsf{Valid}_\Sigma(D)\land\mathsf{Rep}(S,D)\\
&\Longrightarrow\mathsf{Allowed}(D,Q_T(p),E_T(S,p)).
\end{split}
\label{eq:template-target}
\end{equation}
In Example~\ref{ex:running-sqlspec}, $p=(k)$ ranges over $\mathcal{P}=\{(k)\mid -2^{63}\le k\le 2^{63}-1,\ k\in\mathbb{Z}\}$, so a single proof guarantees correctness across every threshold $k$ and every valid database $D$.

In \emph{engine synthesis}, the synthesized engine $E$ does not hardcode specific query templates, table names, or column names. At runtime, $E$ receives a query tree $Q\in\mathcal Q_\Sigma$, where $\mathcal Q_\Sigma$ contains all well-typed SQLSpec query trees over the user-declared schema $\Sigma$ within the supported SQL subset. The fixed formal synthesis target requires $E$ to return a permitted outcome for every such query on every valid database and matching physical storage state:
\begin{equation}
\begin{split}
\forall D,S,\ \forall Q\in\mathcal Q_\Sigma:\quad
&\mathsf{Valid}_\Sigma(D)\land\mathsf{Rep}(S,D)\\
&\Longrightarrow\mathsf{Allowed}(D,Q,E(S,Q)).
\end{split}
\label{eq:engine-target}
\end{equation}
The verified interpreter in Section~\ref{sec:sqlspec-semantics} supports all schemas, databases, and queries within SQLSpec's supported subset, and its correctness guarantee additionally quantifies over the schema $\Sigma$. For continual optimization, GenDB-V instantiates the general engine for the user's declared schema and uses the data, representative workload, and hardware to guide optimization. The engine retains general SQL coverage through verified execution paths, ranging from optimized operators to the reference interpreter, while performance optimization targets the given context.

\noindent\textbf{End-to-end instantiation on imported data.}
Combining the storage targets (Equations~\ref{eq:storage-builder}--\ref{eq:storage-lossless}) with the execution target (Equation~\ref{eq:template-target} or~\ref{eq:engine-target}) establishes end-to-end correctness on imported user data. When GenDB-V imports a valid database $D$ satisfying $\mathsf{Valid}_\Sigma(D)$ and runs the verified storage builder $S=B(D)$, Equation~\ref{eq:storage-builder} establishes $\mathsf{Rep}(S,D)$ and Equation~\ref{eq:storage-lossless} ensures that $S$ uniquely preserves $D$. Instantiating Equation~\ref{eq:template-target} or~\ref{eq:engine-target} with $(D,S)$ then guarantees that running $E_T(S,p)$ for any $p\in\mathcal{P}$, or $E(S,Q)$ for any $Q\in\mathcal Q_\Sigma$, returns a permitted SQLSpec outcome for $D$.

%% file: figures/sqlspec-example-figure.tex
\begin{figure*}[!t]
  \centering
  \input{figures/sqlspec-example}
  \caption{Translating the running example from a SQL template (step~1) to a typed AST (step~2), a SQLSpec query tree $Q_T(k)$ (step~3), and its mathematical semantics (step~4). }
  \label{fig:sqlspec-example}
\end{figure*}
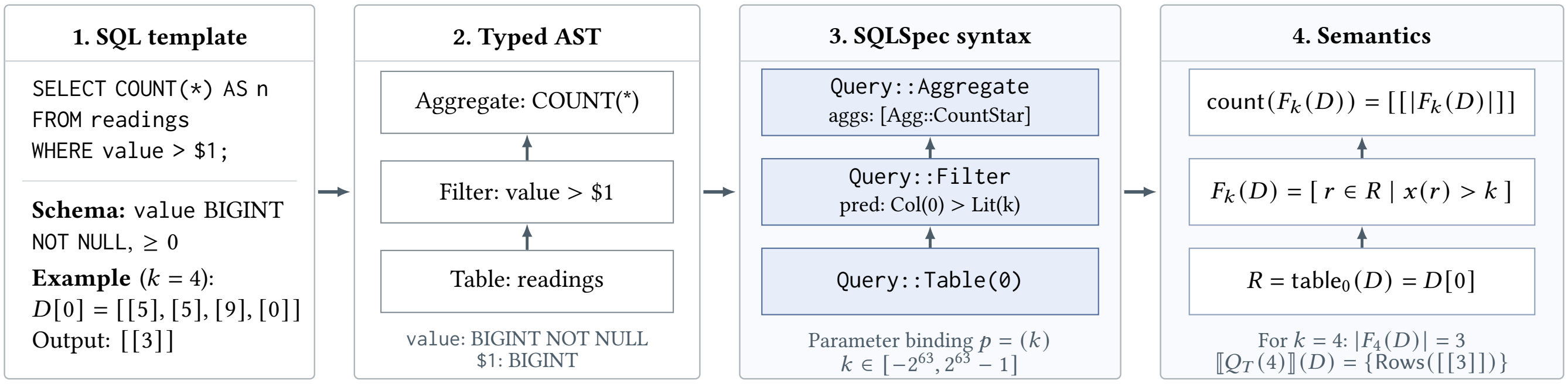

% Operator nodes align horizontally across steps~2--4, with data flowing bottom-to-top from table scan to filter to aggregate.

%% file: figures/sqlspec-example.tex
% Four aligned views of one query body. Mathematical notation is illustrative;
% SQLSpec constructor names are actual names; ownership/sequence wrappers are elided.
\begin{tikzpicture}[x=1mm,y=-0.88mm,>=Latex,font=\small,
  panel/.style={draw=gendbgray!45,line width=.45pt,fill=white,rounded corners=1.5pt},
  astnode/.style={draw=gendbgray!70,fill=white,minimum height=7mm,minimum width=33mm,inner xsep=1.2mm,align=center,font=\small},
  specnode/.style={draw=gendbblue,fill=gendbpaleblue,minimum height=7.5mm,minimum width=38mm,inner xsep=1.2mm,inner ysep=1pt,align=center,font=\ttfamily\small},
  semnode/.style={draw=gendbblue!65,fill=white,minimum height=7.5mm,minimum width=39mm,inner xsep=1.2mm,align=center,font=\small},
  conn/.style={-{Latex[length=1.5mm,width=1.2mm]},draw=gendbgray,line width=.9pt},
  pconn/.style={-{Latex[length=1.5mm,width=1.2mm]},draw=gendbgray,line width=1.1pt}]

% Panels
\path[panel] (0,0) rectangle (35,48);
\path[panel] (39.5,0) rectangle (78.5,48);
\path[panel,fill=gendbpaleblue!30] (83,0) rectangle (126,48);
\path[panel,fill=gendbpaleblue!30] (130.5,0) rectangle (176,48);

% Panel headers
\node[font=\small\bfseries,anchor=north] at (17.5,1.5) {1. SQL template};
\node[font=\small\bfseries,anchor=north] at (59,1.5) {2. Typed AST};
\node[font=\small\bfseries,anchor=north] at (104.5,1.5) {3. SQLSpec syntax};
\node[font=\small\bfseries,anchor=north] at (153.25,1.5) {4. Semantics};
\draw[draw=gendbgray!30,line width=.4pt] (0,7.2)--(35,7.2);
\draw[draw=gendbgray!30,line width=.4pt] (39.5,7.2)--(78.5,7.2);
\draw[draw=gendbgray!30,line width=.4pt] (83,7.2)--(126,7.2);
\draw[draw=gendbgray!30,line width=.4pt] (130.5,7.2)--(176,7.2);

% Panel 1: SQL template
\node[anchor=north west,align=left,text width=31mm,font=\ttfamily\small] at (2,8.3) {
SELECT COUNT(*) AS n\\
FROM readings\\
WHERE value > \$1;
};
\draw[draw=gendbgray!25,line width=.35pt] (2,22.6)--(33,22.6);
\node[anchor=north west,align=left,text width=31mm,font=\small] at (2,23.8) {
\textbf{Schema:} \texttt{value} BIGINT\\
\texttt{NOT NULL}, $\ge 0$\\[0.6mm]
\textbf{Example} ($k=4$):\\
$D[0]=[[5],[5],[9],[0]]$\\
Output: $[[3]]$
};

% Panel 2: Typed AST (arrows point bottom-to-top: Table -> Filter -> Aggregate)
\node[astnode] (agg) at (59,12.5) {Aggregate: COUNT(*)};
\node[astnode] (filter) at (59,24) {Filter: value $>\$1$};
\node[astnode] (scan) at (59,35.5) {Table: readings};
\draw[conn] (scan.north)--(filter.south);
\draw[conn] (filter.north)--(agg.south);
\node[anchor=north,align=center,font=\footnotesize,text=gendbgray] at (59,40.6) {\texttt{value}: BIGINT NOT NULL\\[-1pt]\texttt{\$1}: BIGINT};

% Panel 3: SQLSpec syntax (arrows point bottom-to-top: Table -> Filter -> Aggregate)
\node[specnode] (sagg) at (104.5,12.5) {Query::Aggregate\\[-1pt]{\normalfont\footnotesize aggs: [Agg::CountStar]}};
\node[specnode] (sf) at (104.5,24) {Query::Filter\\[-1pt]{\normalfont\footnotesize pred: Col(0) $>$ Lit(k)}};
\node[specnode] (st) at (104.5,35.5) {Query::Table(0)};
\draw[conn] (st.north)--(sf.south);
\draw[conn] (sf.north)--(sagg.south);
\node[anchor=north,align=center,font=\footnotesize,text=gendbgray] at (104.5,40.6) {Parameter binding $p=(k)$\\[-1pt]$k\in[-2^{63},2^{63}-1]$};

% Panel 4: Semantics (arrows point bottom-to-top: R -> F_k(D) -> [[|F_k(D)|]])
\node[semnode] (magg) at (153.25,12.5) {$\mathsf{count}(F_k(D))=[[|F_k(D)|]]$};
\node[semnode] (mf) at (153.25,24) {$F_k(D)=[\,r\in R\mid x(r)>k\,]$};
\node[semnode] (mt) at (153.25,35.5) {$R=\mathsf{table}_0(D)=D[0]$};
\draw[conn] (mt.north)--(mf.south);
\draw[conn] (mf.north)--(magg.south);
\node[anchor=north,align=center,font=\footnotesize,text=gendbgray] at (153.25,40.6) {For $k=4$: $|F_4(D)|=3$\\[-1pt]$\llbracket Q_T(4)\rrbracket(D)=\{\mathsf{Rows}([[3]])\}$};

% Inter-panel arrows
\draw[pconn] (35.4,24)--(39.1,24);
\draw[pconn] (78.9,24)--(82.6,24);
\draw[pconn] (126.4,24)--(130.1,24);
\end{tikzpicture}

%% file: sections/04-obligations.tex
\section{Joint Code and Proof Synthesis}
\label{sec:joint-synthesis}

The fixed formal targets in Section~\ref{sec:sqlspec} define what outcomes an implementation must produce, but leave open how an agent synthesizes an implementation that is both fast and verifiable. GenDB-V addresses four challenges. The highest specialized performance requires storage layouts and execution algorithms that are designed together (Section~\ref{sec:cooptimization}). Proofs are too expensive to write for every candidate design (Section~\ref{sec:exploration}). A single proof against SQLSpec is too large (Section~\ref{sec:modular-proofs}). Code that is optimized for one database state can become slow when the data changes (Section~\ref{sec:robust-synthesis}).

% through storage--execution co-optimization, performance-first exploration, and modular verification.
% with intermediate contracts.

\subsection{Joint Storage and Query Optimization}
\label{sec:cooptimization}

High-performance analytical engines rely on specialized physical storage layouts together with execution operators tailored to those layouts. Allowing an agent to co-design storage and execution introduces two difficulties: unrestricted storage changes could corrupt data or precompute query answers, and the execution proof must be able to reason formally about custom physical layouts. GenDB-V resolves both difficulties by governing storage customization under formal and policy safeguards, and by exposing physical layout invariants to execution proofs through the predicate $\mathsf{Rep}(S,D)$.

\noindent\textbf{Storage customization and safeguards.}
When storage optimization is enabled, an agent analyzes the schema, data, workload, and hardware to synthesize a custom storage builder $B$, reader functions, and a representation predicate $\mathsf{Rep}(S,D)$. The agent can improve execution through physical designs such as compact encodings, table sorting and partitioning, single-table indexes, and data-skipping metadata. GenDB-V applies two safeguards to preserve data and prevent performance gains from stored query answers. First, GenDB-V automatically invokes Verus to check the storage proofs against the fixed storage targets (Equations~\ref{eq:storage-builder}--\ref{eq:storage-lossless}), establishing builder correctness and losslessness. Second, a configurable storage policy limits storage size and restricts what the agent may store. Our experiments require one complete, lossless copy of the base data and prohibit redundant copies and materialized query computations (Section~\ref{sec:eval-setup}). GenDB-V checks storage sizes programmatically, and we audit the generated code for storage-policy violations to ensure a fair experimental comparison. For ablation studies that isolate query execution optimization from physical storage design, GenDB-V also supports a fixed-storage mode in which agents optimize execution over plain column arrays.

\noindent\textbf{Exposing physical invariants to execution proofs.}
Because the execution targets constrain only the input preconditions and permitted outcomes, agents can design physical execution plans that differ from SQLSpec's recursive operator tree, such as reordered joins, custom hash tables, fused single-pass loops, and parallel execution.
% via GenDB-V's verified \texttt{fork\_join} primitive.
Since an execution proof sees physical storage $S$ only through $\mathsf{Rep}(S,D)$, any physical property that an execution algorithm relies on for correctness must be stated as a mathematical invariant inside $\mathsf{Rep}(S,D)$ and proved by the storage builder $B$. Instead of restricting agents to predefined storage layouts, GenDB-V lets agents specify in $\mathsf{Rep}(S,D)$ both how the physical structures decode to the logical rows of $D$ and what physical properties hold over those structures (such as sort orders, index coverage, or block min/max bounds). In Example~\ref{ex:running-sqlspec}, a plain-array layout states in $\mathsf{Rep}(S,D)$ that array element $v[i]$ equals column~0 of logical row~$i$ ($D[0][i][0]=\mathsf{Int}(v[i])$). If the agent instead partitions the column into blocks $b$ and stores a zone-map maximum $m_b$ for each block, the storage builder proves as part of $\mathsf{Rep}(S,D)$ that every row $r$ in block $b$ satisfies $x(r)\le m_b$. The query proof can then skip any block with $m_b\le k$, because $\mathsf{Rep}(S,D)$ guarantees that no row in $b$ satisfies $x(r)>k$. Agents can similarly exploit schema constraints from $\mathsf{Valid}_\Sigma(D)$: because \texttt{readings.value} has \texttt{CHECK(value >= 0)}, every row in a valid database satisfies $x(r)\ge 0$, allowing a verified fast path to return the table length $|D[0]|$ in $O(1)$ time whenever $k<0$.

\subsection{Performance-First Exploration}
\label{sec:exploration}

Developing a complete proof requires writing loop invariants, arithmetic bounds, and bridge lemmas, which consumes substantial agent reasoning tokens and verifier runtime. If an agent writes proofs before measuring performance, it either wastes proof effort on slow designs or chooses simple execution plans that are easy to prove but slow at runtime. GenDB-V addresses this by decoupling fast empirical performance exploration from proof generation.
% while requiring full verification before any candidate is admitted.

\noindent\textbf{Decoupling performance trials from verification.}
GenDB-V provides an interactive toolset for agents to compile code, build storage, check outputs, time execution, and profile CPU and I/O across parameter bindings and thread counts. During exploration, agents invoke these tools in an unverified diagnostic mode (\texttt{-no-verify}) that compiles and runs candidate code without running the formal verifier. In this mode, an agent can rapidly prototype and compare competing designs, use differential testing against a reference system for fast result feedback, and inspect hardware profiling metrics. Note that the correctness criterion is SQLSpec, not this feedback, because a reference system can be wrong (Section~\ref{sec:eval-robust}). Once the agent identifies the fastest design that meets the user-specified performance target, it freezes the physical algorithm and develops the formal contracts and proofs for that winning design. When the agent submits a candidate implementation, GenDB-V always runs the formal verifier and rejects any code without a complete proof.

\noindent\textbf{Iterative refinement across attempts and versions.}
When a candidate fails verification or misses a performance target, GenDB-V returns structured diagnostic feedback to guide the next iteration, such as storage contracts, query execution time, baseline speed targets, verifier error messages, and optimization notes recorded by agents in prior sessions. In template specialization, GenDB-V evaluates verified submissions across different parameter bindings and thread counts, admitting the fastest verified implementation. In engine synthesis, optimization proceeds iteratively across sequential engine versions starting from the verified reference interpreter (Section~\ref{sec:sqlspec-semantics}). In each iteration, the agent profiles the current best verified engine on a representative workload, identifies bottlenecks, replaces them with faster implementations and develops proofs of correctness, and submits a new complete engine. A candidate engine version replaces the active version only if it passes full verification and improves workload execution time.

\subsection{Modular Contracts and Proofs}
\label{sec:modular-proofs}

\input{figures/proof-example-figure}

Proving an optimized implementation directly against SQLSpec's recursive definitions in a single step makes the proof large and complex for agents to develop and for the verifier to check. In SQLSpec, each operator constructor 
% (e.g., \texttt{Table}, \texttt{Filter}, \texttt{Aggregate}) 
is defined functionally by materializing a full intermediate relation from its child query. However, an optimized physical implementation could fuse operators within a pipeline into tight loops over physical arrays and in-memory state, rather than materializing a full logical relation after every operator. To keep proof steps small and checkable, GenDB-V guides agents to organize proofs into three layers: (1) \emph{storage reader contracts}, (2) \emph{physical loop contracts and step lemmas}, and (3) \emph{inductive bridge lemmas}. Within each layer, agents can further decompose proofs into smaller contracts and helper lemmas as needed.

\noindent\textbf{Three-layer proof decomposition.}
First, at the storage boundary (Layer~1), each storage reader function takes the representation predicate $\mathsf{Rep}(S,D)$ as a precondition (\texttt{requires}) and specifies a postcondition (\texttt{ensures}) describing how the returned physical arrays decode to the logical rows and columns of $D[t]$, along with any physical layout invariants. Second, at the execution loop level (Layer~2), the agent defines mathematical helper functions that describe the computation over physical data. A \emph{loop invariant} (\texttt{invariant}) relates the current program state to these functions. The invariant tracks progress according to the algorithm, such as a scan position, two positions in a nested-loop join, or the hash table built from processed rows. The agent also supplies a \emph{termination measure} (\texttt{decreases}) and helper lemmas for individual steps. Verus checks that the invariant holds at loop entry and that each iteration preserves it. At loop exit, the invariant and exit condition establish the function's postcondition (\texttt{ensures}). Third, at the semantic boundary (Layer~3), the agent writes an \emph{inductive bridge lemma} that combines the Layer~1 storage reader contract with the Layer~2 physical postcondition. The bridge proof connects the physical computation to SQLSpec's evaluation over the logical database $D$, establishing the fixed synthesis target (Equation~\ref{eq:template-target} or~\ref{eq:engine-target}).

\begin{example}
\label{ex:running-proof}
Figure~\ref{fig:proof-example} illustrates this three-layer method on our running COUNT query from Example~\ref{ex:running-sqlspec}. Recall that \texttt{readings} is table~0 ($D[0]$), \texttt{value} is column~0 (\texttt{BIGINT NOT NULL}), and $k$ is a non-null 64-bit integer parameter. In Figure~\ref{fig:proof-example}(a), executable code is on a white background, agent-written proof annotations are in green, and the fixed synthesis target is in blue. The top-level executor $E_T(S,k)$ calls \texttt{read\_\allowbreak values(S)} to obtain the physical 64-bit integer slice $v$ (Layer~1), calls \texttt{count\_\allowbreak above(v, k)} to compute $\mathit{matched}$ in a single fused loop (Layer~2), invokes the bridge lemma \texttt{bridge\_\allowbreak count} (Layer~3), and returns $o=\mathsf{encodeCount}(\mathit{matched})$.

To verify \texttt{count\_\allowbreak above(v, k)} in Layer~2, the agent defines a mathematical prefix-count function $C(v,k,i)$ that counts elements above $k$ in the slice prefix $v[0..i)$:
\begin{equation}
C(v,k,i)=\bigl|\{j\mid 0\le j<i\ \land\ v[j]>k\}\bigr|,
\label{eq:prefix-count}
\end{equation}
and proves a one-step lemma (\texttt{prefix\_\allowbreak step}) showing how $C$ advances from $\mathit{row}$ to $\mathit{row}+1$:
\begin{equation}
C(v,k,\mathit{row}+1)=C(v,k,\mathit{row})+\mathbf{1}[v[\mathit{row}]>k],
\label{eq:prefix-step}
\end{equation}
The indicator $\mathbf{1}[v[\mathit{row}]>k]$ equals $1$ when $v[\mathit{row}]>k$ and $0$ otherwise. As shown in the top row of Figure~\ref{fig:proof-example}(b), the verifier first checks that the loop invariant $\mathit{matched}=C(v,k,\mathit{row})$ holds at loop entry ($\mathit{matched}_0=0=C(v,k,0)$ at $\mathit{row}=0$). Next, the verifier translates the loop body into the logical transition formula $\mathit{matched}'=\mathit{matched}+\mathbf{1}[v[\mathit{row}]>k]$ and $\mathit{row}'=\mathit{row}+1$. Using \texttt{prefix\_\allowbreak step}, the verifier checks that whenever $\mathit{matched}=C(v,k,\mathit{row})$ holds at the start of an iteration, $\mathit{matched}'=\allowbreak C(v,k,\mathit{row})+\mathbf{1}[v[\mathit{row}]>k]=C(v,k,\mathit{row}')$ holds after the iteration (Layer~2a), while $0\le\mathit{matched}\le\mathit{row}<|v|$ guarantees valid array bounds and no counter overflow. When the loop exits at $\mathit{row}=|v|$, the verified invariant establishes the scan postcondition $\mathit{matched}=C(v,k,|v|)=\allowbreak\bigl|\{i\mid 0\le i<|v|,\ v[i]>k\}\bigr|$ (Layer~2b).

In Layer~3, the bridge lemma (\texttt{bridge\_\allowbreak count}) combines the Layer~1 reader contract under $\mathsf{Rep}(S,D)$ (which guarantees $|v|=|D[0]|$ and $x(D[0][i])=v[i]$ for all $0\le i<|v|$) with the Layer~2b scan postcondition ($\mathit{matched}=C(v,k,|v|)$). By induction on $i$, \texttt{bridge\_\allowbreak count} proves that the physical count equals the size of SQLSpec's filtered relation $F_k(D)=[\,r\in D[0]\mid x(r)>k\,]$ from Equation~\ref{eq:example-filter}, giving $\mathit{matched}=C(v,k,|v|)=|F_k(D)|$. Finally, wrapping count $c$:
\begin{equation}
\mathsf{encodeCount}(c)=
\begin{cases}
\mathsf{Rows}([[c]]) & c\le 2^{63}-1,\\
\mathsf{SqlError}(22003) & c>2^{63}-1,
\end{cases}
\label{eq:encode-count}
\end{equation}
matches the permitted outcome set $\llbracket Q_T(k)\rrbracket(D)$ defined in Example~\ref{ex:running-sqlspec}: $\{\,\mathsf{encodeCount}(|F_k(D)|)\,\}$. Here, $\mathsf{Rows}$ denotes a successful query outcome, and $[[c]]$ abbreviates a single row containing the integer value $\mathsf{Int}(c)$, with SQL result type \texttt{BIGINT}. Since $\mathit{matched}=|F_k(D)|$, the returned outcome $o=\mathsf{encodeCount}(\mathit{matched})$ satisfies $o\in\llbracket Q_T(k)\rrbracket(D)$, which holds if and only if $\mathsf{Allowed}(D,Q_T(k),o)$, proving the fixed formal synthesis target across all valid databases $D$ and all 64-bit integer bindings $p=(k)\in\mathcal{P}$.
\end{example}

\noindent\textbf{From template proofs to general-purpose engine proofs.}
A general-purpose engine $E(S,Q)$ receives an arbitrary query tree $Q\in\mathcal Q_\Sigma$ at runtime and must satisfy the engine target (Equation~\ref{eq:engine-target}) for every such query. Queries can nest operators to arbitrary depth, but every query is built from the finite set of constructors in Table~\ref{tab:sqlspec-layers}. The proof therefore follows the structure of the query tree, with one proof per constructor. At a leaf \texttt{Table}$(t)$, a Layer~1 reader contract proves that reading table $t$ from $S$ yields the logical rows of $D[t]$. At an internal operator $q$, each child is a smaller tree, so the proof can assume the postcondition of the recursive calls: each child returns a relation matching $\mathsf{eval}(D,\rho,q_i)$. Layers~2 and~3 then prove that the operator's output matches $\mathsf{eval}(D,\rho,q)$. At the root, the proof of ordering, limit, and offset establishes $\mathsf{Allowed}(D,Q,E(S,Q))$. Because each operator proof relies only on the postconditions of its children, these proofs cover all queries in $\mathcal Q_\Sigma$ and verify the reference interpreter (Section~\ref{sec:sqlspec-semantics}). Agents then optimize the engine within the same framework. They replace materialized intermediate relations with compact representations and faster operators that prove the same per-node postcondition, and they fuse multi-operator subtrees into single-pass kernels or rewrite subplans through proved equivalence lemmas. A runtime check guards each specialized kernel, and evaluation otherwise falls back to the general verified operators.

\subsection{Robust Code Synthesis}
\label{sec:robust-synthesis}

The techniques above produce verified code that is fast on the given database. This result fits read-only analytical workloads, in which the database does not change after loading. Other workloads apply batched updates: tables grow, rows are deleted, and value distributions drift. After an update, the code still returns correct results, because the fixed formal synthesis targets cover every valid database. However, the code can become slow, because the agent optimized the code for one database state. For example, a fast path that requires dense keys no longer applies after rows are deleted. GenDB-V therefore adds a robust code synthesis step for SQL templates, in which an agent first attacks a verified implementation and then generalizes the implementation. The goal is robustness to realistic drift of the data and the query parameters, such as changes in table sizes, row order, key density, and value distributions. Being fast on every valid database is not the goal: some valid databases are not realistic (e.g., a discount of $-10^{12}$), and for them the code remains correct but can be slow.

\noindent\textbf{Co-analysis of schema, data, query, and code.}
Reading the code alone reveals what the code assumes, but not the data properties that the code never considered. The agent therefore analyzes four sources together. The first source is the schema. The schema guarantees only what it declares: types, keys, and constraints. Every other property of the data can change when the data changes. Examples are the value distribution of a column, the number of child rows per foreign-key parent, the row order of a table, and the relative sizes of tables. The agent identifies such properties from the schema and also considers the full domain of each template parameter. The second source is the given database, from which the agent records the current state of each such property. The third source is the SQL template, from which the agent derives the factors that determine execution cost, such as filter selectivities, join cardinalities, and the number and sizes of groups. The fourth source is the code with its proofs, from which the agent extracts the fast paths with their conditions, the fallback paths, and the cost of each path. A \emph{weakness} is a part of the code whose speed depends on a data property that can change, or on a parameter value. Typical weaknesses are a fast path whose condition can fail, a fixed limit or estimate that the data can exceed, a fallback path with a high cost, and parallel execution that becomes unbalanced under skewed data.

\noindent\textbf{Attack and generalization.}
Attack and generalization proceed in a loop. (1) For each weakness, the agent writes a data generator and parameter bindings that expose the weakness. (2) GenDB-V checks that the attack database satisfies the schema and every declared constraint, and that each binding lies in $\mathcal{P}$. GenDB-V then benchmarks the code and the reference engines (DuckDB and Umbra) on the attack database. (3) The agent revises the code and the proofs to remove the weaknesses. (4) GenDB-V verifies the revised version (Section~\ref{sec:verification}) and benchmarks the revised version on the given and attack databases. GenDB-V compares versions by two criteria in order. The first criterion is the number of \emph{performance cliffs}: query calls that do not finish within the user-specified time limit or that are more than an order of magnitude slower than the fastest reference engine. The second criterion is the geometric mean of the execution times relative to the fastest engine. The first criterion has priority because a mean can hide a few very slow query calls. The loop then repeats on the selected version, because each revision introduces new assumptions and new code paths. In the end, the agent reports every weakness that remains.

%% file: figures/proof-example-figure.tex
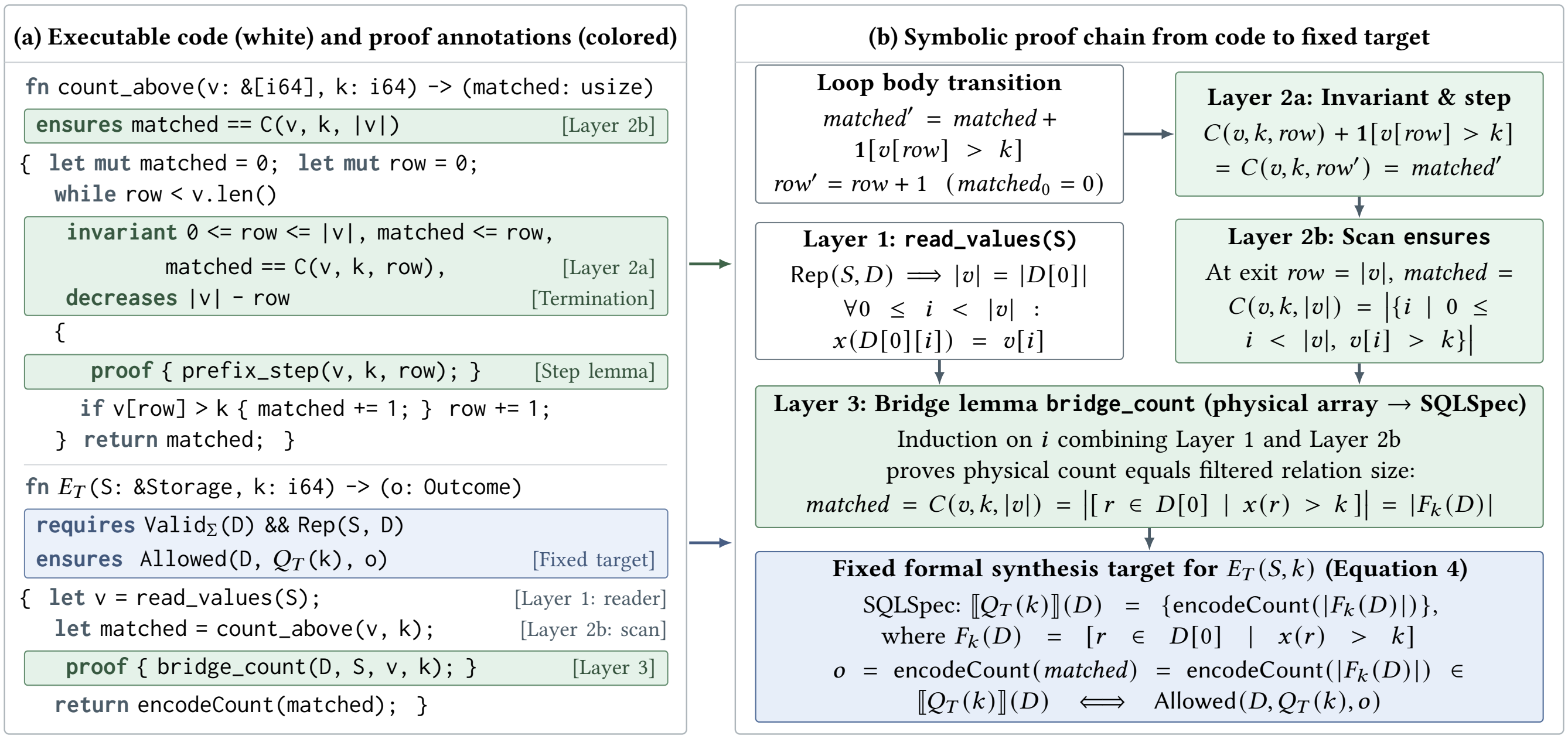
\begin{figure*}[t]
\centering
\begin{tikzpicture}[x=1mm,y=-0.90mm,>=Latex,font=\small,
  panel/.style={draw=gendbgray!45,line width=.45pt,fill=white,rounded corners=1.5pt},
  codebox/.style={anchor=north west,align=left,text width=72.6mm,font=\ttfamily\small,inner sep=0pt},
  proofbox/.style={draw=gendbgreen!85,fill=gendbpalegreen!85,rounded corners=1pt,text width=70.0mm,inner xsep=1.3mm,inner ysep=0.7mm,anchor=north west,align=left,font=\ttfamily\small},
  tbox/.style={draw=gendbblue!85,fill=gendbpaleblue!85,rounded corners=1pt,text width=70.0mm,inner xsep=1.3mm,inner ysep=0.7mm,anchor=north west,align=left,font=\ttfamily\small},
  vcnode/.style={draw=gendbgray!80,fill=white,rounded corners=1.2pt,align=center,text width=39.5mm,minimum height=14mm,inner xsep=1.0mm,inner ysep=0.8mm,font=\small},
  layer2/.style={draw=gendbgreen!85,fill=gendbpalegreen!70,rounded corners=1.2pt,align=center,text width=39.5mm,minimum height=14mm,inner xsep=1.0mm,inner ysep=0.8mm,font=\small},
  bridge/.style={draw=gendbgreen!90,fill=gendbpalegreen!85,rounded corners=1.2pt,align=center,text width=86.5mm,minimum height=14mm,inner xsep=1.2mm,inner ysep=0.8mm,font=\small},
  target/.style={draw=gendbblue,fill=gendbpaleblue,rounded corners=1.2pt,align=center,text width=86.5mm,minimum height=17mm,inner xsep=1.2mm,inner ysep=0.8mm,font=\small},
  conn/.style={-{Latex[length=1.5mm,width=1.2mm]},draw=gendbgray,line width=.9pt},
  pconn/.style={-{Latex[length=1.5mm,width=1.2mm]},draw=gendbgreen!80!black,line width=.9pt},
  bconn/.style={-{Latex[length=1.5mm,width=1.2mm]},draw=gendbblue!85!black,line width=.9pt}]

% Outer panels: Panel (a) 0..77, 5.5mm gap 77..82.5, Panel (b) 82.5..176, height 91.5 units
\path[panel] (0,0) rectangle (77,91.5);
\path[panel] (82.5,0) rectangle (176,91.5);

% Panel headers
\node[font=\small\bfseries,anchor=north] at (38.5,1.5) {(a) Executable code (white) and proof annotations (colored)};
\node[font=\small\bfseries,anchor=north] at (129.25,1.5) {(b) Symbolic proof chain from code to fixed target};
\draw[draw=gendbgray!30,line width=.4pt] (0,7.2)--(77,7.2);
\draw[draw=gendbgray!30,line width=.4pt] (82.5,7.2)--(176,7.2);

% ---------------- Panel (a): All boxes aligned at x=2.2..74.8 ----------------
\node[codebox] at (2.2,9.0) {\textbf{\color{gendbgray}fn} count\_above(v: \&[i64], k: i64) -> (matched: usize)};

\node[proofbox] (ensbox) at (2.2,13.0) {%
  \textbf{\color{gendbgreen!70!black}ensures} matched == C(v, k, |v|)\hfill{\normalfont\footnotesize\color{gendbgreen!65!black}[Layer 2b]}%
};

\node[codebox] at (2.2,18.6) {\{\ \ \textbf{\color{gendbgray}let mut} matched = 0;\ \ \textbf{\color{gendbgray}let mut} row = 0;};
\node[codebox] at (2.2,22.5) {\hspace*{1.2em}\textbf{\color{gendbgray}while} row < v.len()};

\node[proofbox] (invbox) at (2.2,26.5) {%
  \hspace*{1.2em}\textbf{\color{gendbgreen!70!black}invariant} 0 <= row <= |v|, matched <= row,\\[1pt]
  \hspace*{5.2em}matched == C(v, k, row),\hfill{\normalfont\footnotesize\color{gendbgreen!65!black}[Layer 2a]}\\
  \hspace*{1.2em}\textbf{\color{gendbgreen!70!black}decreases} |v| - row\hfill{\normalfont\footnotesize\color{gendbgreen!65!black}[Termination]}%
};

\node[codebox] at (2.2,39.8) {\hspace*{1.2em}\{};

\node[proofbox] (stepbox) at (2.2,43.8) {%
  \hspace*{2.2em}\textbf{\color{gendbgreen!70!black}proof} \{ prefix\_step(v, k, row); \}\hfill{\normalfont\footnotesize\color{gendbgreen!65!black}[Step lemma]}%
};

\node[codebox] at (2.2,49.4) {\hspace*{2.2em}\textbf{\color{gendbgray}if} v[row] > k \{ matched += 1; \}\ \ row += 1;};
\node[codebox] at (2.2,53.3) {\hspace*{1.2em}\}\ \ \textbf{\color{gendbgray}return} matched;\ \ \}};

\draw[draw=gendbgray!25,line width=.35pt] (2.2,57.7)--(74.8,57.7);

% Top-level executor E_T(S, k)
\node[codebox] at (2.2,59.2) {\textbf{\color{gendbgray}fn} $E_T$(S: \&Storage, k: i64) -> (o: Outcome)};
\node[tbox] (tensbox) at (2.2,63.2) {%
  \textbf{\color{gendbblue!80!black}requires} Valid$_\Sigma$(D) \&\& Rep(S, D)\\[1pt]
  \textbf{\color{gendbblue!80!black}ensures} \ Allowed(D, $Q_T$(k), o)\hfill{\normalfont\footnotesize\color{gendbblue!80!black}[Fixed target]}%
};
\node[codebox] at (2.2,73.1) {\{\ \ \textbf{\color{gendbgray}let} v = read\_values(S);\hfill{\normalfont\footnotesize\color{gendbgray}[Layer 1: reader]}};
\node[codebox] at (2.2,77.0) {\hspace*{1.2em}\textbf{\color{gendbgray}let} matched = count\_above(v, k);\hfill{\normalfont\footnotesize\color{gendbgray}[Layer 2b: scan]}};
\node[proofbox] (brbox) at (2.2,81.0) {%
  \hspace*{1.2em}\textbf{\color{gendbgreen!70!black}proof} \{ bridge\_count(D, S, v, k); \}\hfill{\normalfont\footnotesize\color{gendbgreen!65!black}[Layer 3]}%
};
\node[codebox] at (2.2,86.6) {\hspace*{1.2em}\textbf{\color{gendbgray}return} encodeCount(matched);\ \ \}};

% ---------------- Panel (b): All boxes aligned at x=84.8..173.7 ----------------
% Row 1 (y=16.2): Code transition formula (left, center 105.55) ---> Layer 2a Invariant & prefix_step (right, center 152.95)
\node[vcnode] (l2vc) at (105.55,16.2) {\textbf{Loop body transition}\\[1.5pt]
$\mathit{matched}'\!=\!\mathit{matched}\!+\!\mathbf{1}[v[\mathit{row}]\!>\!k]$\\[1pt]
$\mathit{row}'=\mathit{row}+1\ \ (\mathit{matched}_0=0)$};

\node[layer2] (l2loop) at (152.95,16.2) {\textbf{Layer 2a: Invariant \& step}\\[1.5pt]
$C(v,k,\mathit{row})+\mathbf{1}[v[\mathit{row}]>k]$\\[1pt]
$=C(v,k,\mathit{row}')=\mathit{matched}'$};

\draw[conn] (l2vc.east) -- (l2loop.west);

% Row 2 (y=35.9): Layer 1 reader contract (left) and Layer 2b Scan postcondition (right)
\node[vcnode] (l1read) at (105.55,35.9) {\textbf{Layer 1: \texttt{read\_values(S)}}\\[1.5pt]
$\mathsf{Rep}(S,D)\Longrightarrow |v|=|D[0]|$\\[1pt]
$\forall 0\le i<|v|:\ x(D[0][i])=v[i]$};

\node[layer2] (l2post) at (152.95,35.9) {\textbf{Layer 2b: Scan \texttt{ensures}}\\[1.5pt]
At exit $\mathit{row}=|v|$, $\mathit{matched}=$\\[1pt]
$C(v,k,|v|)=\bigl|\{i\mid 0\le i<|v|,\ v[i]>k\}\bigr|$};

\draw[conn] (l2loop.south) -- (l2post.north);

% Row 3 (y=56.7): Layer 3 Inductive bridge lemma
\node[bridge] (l3bridge) at (129.25,56.7) {\textbf{Layer 3: Bridge lemma \texttt{bridge\_count} (physical array $\rightarrow$ SQLSpec)}\\[1.5pt]
Induction on $i$ combining Layer~1 and Layer~2b proves physical count equals filtered relation size:\\[1pt]
$\mathit{matched} = C(v,k,|v|) = \bigl|[\,r\in D[0]\mid x(r)>k\,]\bigr| = |F_k(D)|$};

\draw[conn] (l1read.south) -- (l1read.south |- l3bridge.north);
\draw[conn] (l2post.south) -- (l2post.south |- l3bridge.north);

% Row 4 (y=79.3): Fixed formal synthesis target
\node[target] (ftarget) at (129.25,79.3) {\textbf{Fixed formal synthesis target for $E_T(S,k)$ (Equation~\ref{eq:template-target})}\\[1.5pt]
SQLSpec: $\llbracket Q_T(k)\rrbracket(D)=\{\mathsf{encodeCount}(|F_k(D)|)\}$, where $F_k(D)=[r\in D[0]\mid x(r)>k]$\\[1pt]
$o=\mathsf{encodeCount}(\mathit{matched})=\mathsf{encodeCount}(|F_k(D)|)\in\llbracket Q_T(k)\rrbracket(D)\Longleftrightarrow\mathsf{Allowed}(D,Q_T(k),o)$};

\draw[conn] (l3bridge.south) -- (ftarget.north);

% Clean horizontal arrows across the gap connecting Panel (a) to Panel (b)
\draw[pconn] (77.3,32.5) -- (82.2,32.5);
\draw[bconn] (77.3,67.5) -- (82.2,67.5);

\end{tikzpicture}
\caption{Modular code and proof synthesis for Example~\ref{ex:running-proof}.}
\label{fig:proof-example}
\end{figure*}

 % (a) Code (white), agent-written proof annotations (green), and the fixed synthesis target (blue). (b) Reader contracts, loop proofs, and an inductive bridge connect the code to SQLSpec.

%% file: sections/05-synthesis.tex
\section{Independent Verification}
\label{sec:verification}

% After an agent synthesizes candidate code and proofs (Section~\ref{sec:joint-synthesis}), GenDB-V independently verifies that the submission satisfies the fixed formal synthesis targets (Section~\ref{sec:sqlspec}) without trusting any agent-generated code or proof. 

% This section describes how GenDB-V checks and compiles the complete proof chain (Section~\ref{sec:checking}) and summarizes the resulting guarantees and trusted foundation (Section~\ref{sec:guarantees}).

\subsection{Verification and Compilation Pipeline}
\label{sec:checking}

When an agent submits candidate code and proofs, GenDB-V checks that it proves the exact synthesis target, does not skip any proof steps, and produces the executable that matches the verified code. 
% GenDB-V enforces these requirements through a three-step pipeline.

\noindent\textbf{Binding agent code to the fixed targets.}
To prevent reward hacking, GenDB-V runs verification in an isolated environment. In this environment, all specification files (SQLSpec, the schema and constraints $\mathsf{Valid}_\Sigma(D)$, parameter domains $\mathcal{P}$, and the admitted storage specification $\mathsf{Rep}(S,D)$) and a top-level entry-point function are fixed and read-only. The fixed entry point first validates runtime inputs, rejecting out-of-range parameters, unsupported queries ($Q\notin\mathcal Q_\Sigma$), or invalid thread counts. It then calls the agent's implementation function and enforces the fixed formal synthesis target (Equations~\ref{eq:storage-builder}--\ref{eq:engine-target}) as its required postcondition (\texttt{ensures}). Because the agent can only write the code and helper proofs called by this fixed entry point, it cannot weaken the target, add extra input assumptions, or prove an unused function.

\noindent\textbf{Checking code and proofs with Verus.}
GenDB-V builds its verifier on Verus~\cite{lattuada2023verus,verusguide}, a formal verification tool that extends Rust with specification and proof annotations. Agents in GenDB-V write storage and query implementations in a safe subset of Rust that Verus can translate into logical formulas. For each function, Verus translates the executable Rust code, preconditions (\texttt{requires}), postconditions (\texttt{ensures}), loop invariants, and proof lemmas into logical verification conditions. It then uses the Z3 SMT solver to check that every function satisfies its contracts and that the top-level contracts establish the fixed formal synthesis target. To prevent agents from bypassing these checks, GenDB-V deterministically rejects unsafe language features (such as \texttt{unsafe} pointers and external libraries) and runs Verus across the entire program with all proof-skipping annotations (such as \texttt{assume} and \texttt{admit}) disabled.

\noindent\textbf{Single-pass verification and compilation.}
To ensure that the binary executed at runtime matches the verified source code, GenDB-V runs Verus in a single verify-and-compile pass (\texttt{--compile}). Once all proofs check, the compiler erases all specification functions, loop invariants, and proof lemmas, and compiles the remaining executable Rust code into a shared library. At query serving time, GenDB-V loads this verified library and its verified storage files directly, executing queries at native compiled speed with zero runtime overhead from specifications or proofs.

\subsection{Verified Guarantees and Trusted Foundation}
\label{sec:guarantees}
\label{sec:trusted-base}

% \begin{table}[t]
% \centering\small
% \caption{Verified Guarantees from GenDB-V.}
% \label{tab:guarantees}
% \setlength{\tabcolsep}{3pt}
% \begin{tabular*}{\columnwidth}{@{}l@{\extracolsep{\fill}}p{.71\columnwidth}@{}}
% \toprule
% Guarantee & Checked requirement \\
% \midrule
% SQL correctness & Returned rows or SQL errors satisfy SQLSpec's permitted outcomes $\llbracket Q\rrbracket(D)$. \\
% Storage soundness & For builder output $S=B(D)$, builders prove $\mathsf{Rep}(S,D)$ (Eq.~\ref{eq:storage-builder}) and, separately, losslessness (Eq.~\ref{eq:storage-lossless}). Readers prove contracts using the same $\mathsf{Rep}(S,D)$. \\
% Memory safety & Safe Rust ownership and borrowing hold; array indexing and slice operations satisfy proved bounds. \\
% Arithmetic safety & Machine arithmetic and casts satisfy range bounds; SQL overflow raises the specified SQLSpec error. \\
% Parallel execution & Worker tasks are race-free and verified; partitions cover the input and merges preserve results. \\
% Termination & Decreasing measures on all loops and recursive functions and call-graph checks prove termination. \\
% \bottomrule
% \end{tabular*}
% \end{table}

\noindent\textbf{Semantic and program safety guarantees.}
First, for semantic correctness, Verus proves that storage builders and readers preserve all stored data (Equations~\ref{eq:storage-builder}--\ref{eq:storage-lossless}) and that query execution satisfies the permitted outcome targets (Equations~\ref{eq:template-target} or \ref{eq:engine-target}). 
% Readers use the same representation predicate established by the builders. 
Second, for program safety, safe Rust and Verus checks guarantee that execution is free of memory corruption, out-of-bounds array indexing, and silent arithmetic overflow. SQL overflow raises the error specified by SQLSpec. For parallel execution, the current implementation supports fork--join parallelism with shared read-only storage and worker-owned mutable state, providing verified correctness and strong performance in our experiments. Proofs establish complete input coverage and correct result merging. Extending support to shared mutable structures and more parallelization strategies is future work. Finally, Verus checks a strictly decreasing measure on every loop and recursive function, guaranteeing termination.

\noindent\textbf{Trusted foundation.}
% GenDB-V establishes these guarantees without trusting any agent-generated code, intermediate contracts, or proof lemmas. Instead, verification relies on four explicit trusted components. First, we trust that SQLSpec and the deterministic SQL-to-SQLSpec translator accurately capture the intended PostgreSQL semantics, and that input databases satisfy the declared schema constraints in $\mathsf{Valid}_\Sigma(D)$. Second, we trust data loading and storage I/O to pass the input tables $D$ faithfully to the storage builder $B(D)$ and provide the resulting storage $S$ to query execution. Third, we trust the verification and compilation toolchain (Verus, the Z3 SMT solver, and the Rust compiler) to be sound. Fourth, we assume that the operating system and hardware provide sufficient memory, stack space, and fair thread scheduling.
GenDB-V establishes the guarantees without trusting agent-generated code, contracts, or proofs. We trust GenDB-V's SQLSpec and its deterministic translation and target construction. We also trust the external verification and execution infrastructure: Verus, Z3, the Rust compiler and runtime, the operating system, and hardware. The proofs apply to databases satisfying the declared schema constraints. Termination assumes that the environment provides sufficient memory and stack space, and that the operating system does not indefinitely starve runnable threads.

%% file: sections/06-evaluation.tex
\section{Evaluation}
\label{sec:evaluation}

% Evidence: docs/records/RESULTS_INDEX_2026-10-29.md and the underlying FT, E4, E5,
% SEC, SYNX, RUN/RUN2 reports. Detailed measurement and audit notes remain in the
% experiment records; figures/data records the plotted data and accounting scope.

\subsection{Experimental Setup}
\label{sec:eval-setup}

\noindent\textbf{Hardware and models.}
All experiments use a server with two 16-core Intel Xeon Gold 5218 processors (32 physical cores, 64 hardware threads) and 384\,GB of RAM.
GenDB-V and the LLM-based baseline use Claude Opus~5.5 with high effort.

\noindent\textbf{Workloads.}
We use all 22 TPC-H templates at SF100~\cite{tpch} and all 6 SEC-EDGAR templates at SF3~\cite{lao2026gendb}.
TPC-H is well studied, and LLMs may have encountered its queries and optimization techniques during training.
SEC-EDGAR is a new benchmark introduced in GenDB~\cite{lao2026gendb} to reduce training-data leakage.

\noindent\textbf{Databases.}
% Bindings: given databases 3 per template; held-out 5 per template per database (JCC-H: 5 normal and 5 skewed); parameter drift 20 per template at each of 4 levels (independent/drift/DATABASES.md, results/SUMMARY.md); checked 2026-11-29.
We use three kinds of databases.
The \emph{given databases} are TPC-H SF100 and SEC-EDGAR SF3, for which agents synthesize code.
Each template has three parameter bindings on its given database, generated by the official \texttt{qgen} tool for TPC-H and sampled from the benchmark parameter domains for SEC-EDGAR.
The robustness evaluation (Section~\ref{sec:eval-robust}) uses two further kinds.
The \emph{attack databases} are 86 databases that the agents built to expose weaknesses of their own code, each with its own parameter bindings.
The \emph{held-out databases} are 29 databases that we created independently of the synthesis, each with five bindings per template.
Their creation used no synthesized code and no attack database, and no agent had access to the held-out databases, their generators, or their parameter bindings.
The held-out databases are TPC-H SF10, TPC-H SF10 after eight pairs of the official refresh functions, JCC-H~\cite{boncz2017jcch} at SF10 with its normal and skewed parameter bindings, and 26 databases with systematic drift of TPC-H SF10 in six data properties.
Of these, 17 databases change one property, and 9 mixed databases change all properties except the size.
On TPC-H SF10, we also use parameter values beyond the ranges of \texttt{qgen}, with 20 bindings per template at each level.

\noindent\textbf{GenDB-V settings.}
We evaluate three settings.
In \emph{template synthesis over plain arrays}, agents synthesize query code over fixed, plain typed columns in the original row order, without sorting, compression, indexes, or skipping metadata.
In \emph{joint storage and query optimization}, agents co-design physical storage layouts and query code.
For template synthesis, we target at least $5\times$ speedup over the faster of DuckDB and Umbra across parameter bindings and thread counts with at most 5 submissions.
In \emph{general-purpose engine synthesis}, agents iteratively optimize the verified interpreter to reduce total workload time relative to DuckDB and Umbra.
We apply robust code synthesis to the plain-array TPC-H implementations, with a cost limit of \$100 per session and attack databases of at most one tenth of the size of the given database.
All other code is optimized for one given database state and workload, the setting of read-only analytical workloads.

\noindent\textbf{Baselines.}
We compare with DuckDB 1.5.6~\cite{raasveldt2019duckdb} and Umbra 26.08 (the version with ``latest'' tag)~\cite{neumann2020umbra}, using Umbra's optimized compilation mode and each system's default physical design.
For unverified synthesis, we use Bespoke OLAP~\cite{wehrstein2026bespoke} through its maintained implementation~\cite{synnodb}, with the same model and settings as GenDB-V.

\noindent\textbf{Measurements.}
A \emph{run} is one template with one parameter binding on one database at one thread count, and every system executes every run.
On the given databases, we measure at 1, 32, and 64 threads.
Each run consists of one cold execution followed by three hot executions, and we report the median of the three hot executions.
We average each template's time across its three bindings, and we report total workload time and geometric-mean speedups across templates.
For LLM-based methods, we also measure up to three unseen bindings per template to check for overfitting (Section~\ref{sec:eval-unverified}).
On the attack databases, we measure at 1, 32, and 64 threads with a time limit of 60\,s.
On the held-out databases, we measure at 32 threads, and each run has one time limit for all systems: 100 times the time of the faster of DuckDB and Umbra, with a minimum of 10\,s, and a maximum of 600\,s to control the experiment cost.
% Total workload time is the sum of the per-template average times across all templates, and summary speedups use the geometric mean of the per-template speedups.

\subsection{Verified and Robust Template Synthesis}
\label{sec:eval-robust}

GenDB-V synthesizes formally verified and robust implementations for all 22 TPC-H templates over plain arrays. Template synthesis first produces verified code that is optimized for the given database, TPC-H SF100. Robust code synthesis (Section~\ref{sec:robust-synthesis}) then generalizes this code beyond the given database. We call the code before this step the \emph{initial} code and the code after this step the \emph{robust} code, and we compare both with DuckDB and Umbra. Table~\ref{tab:eval-robust} reports the results. The total time is the sum of the execution times of all runs, where a run that is stopped at its time limit counts with that limit. A run is \emph{very slow} when a system takes more than one second and more than 10 times as long as the fastest system on the same run, or when the run is stopped at its time limit. The worst slowdown is the largest ratio between the time of a system and the time of the fastest system on the same run.

\input{figures/eval-robustness-table}

\noindent\textbf{Given database.}
On TPC-H SF100, the robust code needs 182\,s in total, compared with 1,310\,s for DuckDB and 567\,s for Umbra (Table~\ref{tab:eval-robust}). Neither the initial code nor the robust code of GenDB-V has a very slow run. The robust code outperforms DuckDB on 64 of the 66 template--thread combinations and Umbra on 54 of 66, with geometric-mean speedups of $5.69\times$ and $2.22\times$. The generalization does cost performance on the given database, but the cost is small. In the geometric mean over the templates, the robust code needs 8\% more time than the initial code.

\noindent\textbf{Attack and held-out databases.}
The initial code does not generalize: 595 of its 750 runs on the attack databases and 1,215 of its 5,060 runs on the held-out databases are very slow. The robust code has no very slow run and the lowest total time of all systems. The attack databases that the agents built against their own code therefore also cover drift that independent generators produce.

\noindent\textbf{Worst cases.}
The worst slowdowns of the robust code, 306 and 49 times, occur on two runs with an empty result. The robust code must scan its plain arrays, which have no skipping metadata, while Umbra skips all data with its metadata. The robust code is not fast on every valid database. For example, a line item with a discount of $-10^{12}$ sends Q15 to a serial path with quadratic cost, and part keys near $2^{63}$ do the same for Q16. Such databases satisfy the schema but lie outside realistic drift. On the held-out databases, all 26 very slow runs of DuckDB and 11 of the 22 very slow runs of Umbra belong to Q21. The extreme case is Q21 with the five skewed bindings of JCC-H. Umbra needs 582 to 609\,s for each binding, and DuckDB exceeds our limit of 250\,GB of temporary files. These five runs account for about 3,000\,s of the total time of each engine. The robust code answers them in 54 to 64\,ms. JCC-H also exposes a wrong result in Umbra 26.08. For the skewed bindings of Q2, direct execution in Umbra returns an empty result, whereas its prepared-statement execution, DuckDB, and GenDB-V return the same non-empty result. The cause is that a \texttt{MIN}/\texttt{MAX} aggregate in Umbra's group-join operator returns 0 for groups whose values are all positive/negative. Umbra 26.10 and later versions return the correct result.

\input{figures/robustness-figure}

% SF1 cause: reader check of the initial code of Q10 and Q13 (GenDB-codex .lane-tmp/general-engine/EV1/SF1-CAUSE.md, 2026-11-29): bucket capacity assumes an even spread of customers over a rounded-up number of buckets; derived from the code and TPC-H row counts, not observed in a run (Q13's computed serial cost of about 7.5 s matches the measured 7.2-8.1 s).
% Updates and JCC-H panels (figures/data/robustness-groups-time.csv, tools/extract-heldout-groups.py, 2026-11-30; sums checked against the table): mean per run after 8 refresh pairs: DuckDB 0.135, Umbra 0.035, initial 0.507, robust 0.025 s; JCC-H normal: 0.109, 0.029, 0.931, 0.024 s; JCC-H skewed: 27.5, 28.5, 4.25, 0.062 s. Robust fastest at 20 of the 26 changed points (not at row order S/M/L, parameters S, mixed S/M); 17 of the 23 systematic-drift points.
% Initial code, mean time per run (figures/data/robustness-levels-time.csv, checked 2026-11-29): unchanged 0.024 s; first level: row order 7.33, key density 4.85, fan-out 0.94, domain size 0.50, mixed 7.43 s; largest level: 8.04, 4.89, 0.93, 0.67, 10.06 s. Value skew 0.025 / 0.024 / 0.477 s; parameters 0.018 / 0.019 / 0.030 s. SF1: 0.80 s (Q10: 5 runs stopped at 10 s; Q13: 5 runs of 7.2-8.1 s), SF30: 0.063 s.
\noindent\textbf{Effect of each kind of drift.}
Figure~\ref{fig:eval-robust} shows the mean execution time per run on the held-out databases. The initial code degrades abruptly under most kinds of drift. At the first drift level of row order, key density, fan-out, and domain size, and on the mixed databases, its mean time per run grows from 24\,ms to between 0.5\,s and 7.4\,s, and larger drift adds little. This step shape is consistent with fast paths that stop applying once their conditions fail, and Section~\ref{sec:eval-cases} shows an example for Q21. A smaller database also slows the initial code down: at SF1, all runs of Q10 and Q13 are very slow, because both implementations divide the customers into buckets of a fixed capacity, and a bucket that overflows sends the query to a serial fallback path. The robust code changes little with the amount of drift and is the fastest system at 20 of the 26 changed points. After updates and on JCC-H, the initial code needs 0.5\,s to 4.3\,s per run, and the robust code needs at most 0.06\,s. Under fan-out drift, the times of DuckDB and Umbra grow by factors of 13 and 4, while the time of the robust code does not grow. Row order is the main exception: once rows are no longer stored in key order, the robust code needs about 1.4 times the time of Umbra. On the mixed databases, the robust code needs 1.2 times the time of Umbra at levels S and M and is 10 times faster at level L, where single runs of Umbra take up to 108\,s.

\subsection{Joint Optimization and Unfamiliar Queries}
\label{sec:eval-joint}
\label{sec:eval-sec}

% Numbers: figures/data/template-visible-mean.csv (FT window), rechecked 2026-11-29: joint 51.6x / 22.0x, faster than both baselines on 66 of 66; joint vs initial plain-array code 9.4x.
\noindent\textbf{Joint storage and query optimization.}
The storage that agents design must follow a fixed policy.
The storage must preserve all base data losslessly.
We cap storage at the uncompressed source-file size: 114.64\,GB for TPC-H's pipe-delimited files generated by \texttt{dbgen}, and 5.44\,GB for SEC-EDGAR's cleaned, tab-separated financial-statement files.
Agents may optimize layouts, row orders, and encodings, and add skipping metadata or single-table indexes containing only search keys and row positions.
We prohibit redundant base-data copies, cross-table row locators, stored derived values, materialized joins, precomputed aggregates, and cached query answers.
Our audit of the generated storage and query implementations found no policy violations.
With joint optimization, GenDB-V outperforms both baselines on every TPC-H template at all three thread counts.
The geometric-mean speedups across templates and thread counts are $51.6\times$ over DuckDB and $22.0\times$ over Umbra, a further $9.4\times$ over the initial plain-array code.
At 32 threads, the total TPC-H workload time is 21.3\,s for DuckDB, 10.2\,s for Umbra, 4.5\,s for the initial plain-array code, and 0.84\,s with joint optimization.
This code and its storage are specialized to the given database state.

\noindent\textbf{Unfamiliar benchmark.}
% Numbers: figures/data/template-visible-mean.csv, rechecked 2026-11-29: plain 3.99x / 1.84x (faster on 18 and 15 of 18), joint 61.3x / 28.3x (18 and 18 of 18), joint vs plain 15.4x.
SEC-EDGAR tests whether the synthesis depends on a benchmark that LLMs know from their training data.
GenDB-V synthesizes formally verified implementations for all 6 SEC-EDGAR templates under both storage settings.
Over plain arrays, the code outperforms DuckDB on all 18 template--thread combinations and Umbra on 15 of 18, with geometric-mean speedups of $3.99\times$ and $1.84\times$.
With joint optimization, the code outperforms both baselines on all 18 combinations, with geometric-mean speedups of $61.3\times$ and $28.3\times$, a further $15.4\times$ over plain arrays.
The general-purpose engine is also faster than both baselines in total workload time on SEC-EDGAR (Section~\ref{sec:eval-engine}).
The gains on both workloads show that the approach extends beyond TPC-H query patterns that are familiar to LLMs.

\subsection{Verified General-Purpose Engine Synthesis}
\label{sec:eval-engine}

% Engine = the final version (v020) of the one-engine run G3; final evaluation of 2026-11-05 (GenDB-codex .lane-tmp/general-engine/G3/run/engine/final/v020-{tpch-sf100,sec-sf3}/final-evaluation.json): 1 warm + 3 hot calls, median, binding v0 only, no incident.
% Totals in s at 1/32/64 threads (22 TPC-H templates; inventory EN1/ENGINE-RESULTS.md, checked 2026-11-29): engine 93.645 / 8.352 / 7.392, DuckDB 375.456 / 21.566 / 16.353, Umbra 165.849 / 9.987 / 11.647. SEC-EDGAR: engine 5.568 / 0.708 / 0.657, DuckDB 27.926 / 1.830 / 1.593, Umbra 11.411 / 0.772 / 1.131. The DuckDB and Umbra times are saved times of the same binding from other days.
% Per template (EN1/pertemplate.py): faster than DuckDB on 22/22/20 of 22 TPC-H templates and 6/6/6 of 6 SEC-EDGAR templates; faster than Umbra on 16/10/15 and 5/3/6.
% Speedups at 1/32/64 threads: TPC-H over DuckDB 4.01 / 2.58 / 2.21, over Umbra 1.77 / 1.20 / 1.58; SEC-EDGAR over DuckDB 5.02 / 2.59 / 2.42, over Umbra 2.05 / 1.09 / 1.72. The optimization workload also held 15 additional TPC-H queries (set A); the owner decided on 2026-11-29 not to mention them.
% Lineage: run 2 (interpreter to v016) -> continual evolution (15 further TPC-H queries, set A) -> SEC-EDGAR stage -> one-engine run G3. The same code serves both schemas from G3 v004 on; Verus verifies 7,182 items with 0 errors per schema.
% The earlier text reported v016 of run 2 from a development race (1 hot call); under the 3-hot-call protocol v016 is slower than Umbra at 32 threads (0.93x).
GenDB-V produces one verified engine that executes any well-typed query expressible in SQLSpec.
The same engine code serves TPC-H and SEC-EDGAR.
Agents developed the engine iteratively from the verified reference interpreter, which completes only 2 of 22 TPC-H queries within 60\,s at SF1.
We measure the final engine with one parameter binding per template.
In total workload time at 1, 32, and 64 threads, the engine is $2.21$--$4.01\times$ faster than DuckDB and $1.20$--$1.77\times$ faster than Umbra on TPC-H SF100, and $2.42$--$5.02\times$ and $1.09$--$2.05\times$ faster on SEC-EDGAR SF3.
The engine is faster than DuckDB on 64 of the 66 TPC-H template--thread combinations and faster than Umbra on 41 of 66.

% Plan choices need no proof: review of the final engine (R1): every plan choice, threshold and sample is outside the proof; the invariants do not mention the plan or the join order.
The engine combines compact column storage, formally verified query rewrites, join planning at query time, and specialized parallel operators.
Agents optimize these components for the target workload, but every component remains correct for all queries within SQLSpec's fragment.
The engine applies the verified rewrite rules to incoming queries at runtime without new proofs, and plan choices such as the join order need no proof.
General operators and the verified interpreter handle queries outside the applicability conditions of the specialized operators.
The engine is optimized for the given databases and workloads, and applying robust code synthesis to the general-purpose engine is future work.

\subsection{Comparison with Unverified Synthesis}
\label{sec:eval-unverified}

\noindent\textbf{Performance.}
On the given databases, we compare the jointly optimized GenDB-V code and Bespoke OLAP by the geometric mean of the per-template execution times, on the parameter values that were visible during synthesis and on unseen values.
GenDB-V and Bespoke OLAP achieve comparable performance at 1 thread and both substantially outperform DuckDB and Umbra, while GenDB-V is faster at larger thread counts.
Across both workloads and both sets of parameter values, GenDB-V is $1.08$--$1.28\times$ faster than Bespoke OLAP at 1 thread, $1.92$--$4.87\times$ at 32 threads, and $2.75$--$6.79\times$ at 64 threads.
On unseen parameter values, the geometric mean of each system is at most 11.3\% higher than on the visible values, which shows that the generated code generalizes beyond the bindings used during synthesis.
Bespoke OLAP's agent is prompted to use the configured thread count, but Bespoke OLAP is slower at 64 than at 32 threads because of thread-management costs and limited additional parallel speedup.
GenDB-V follows a different design: the configured thread count is an upper bound, and agents choose whether to parallelize and how many threads to use.
For example, the generated Q2 implementation remains serial even when 64 threads are available.

\input{figures/generalization-figure}

\noindent\textbf{Robustness.}
GenDB-V (initial) and Bespoke OLAP both synthesize code that is specialized for one database. Figure~\ref{fig:eval-generalization} shows how often this specialized code remains faster than DuckDB, the general-purpose engine that a system can fall back to. Bespoke OLAP's generated engine rejects the data of all 86 attack databases and of 19 of the 29 held-out databases, and its router then answers the queries through DuckDB. Its specialized code is therefore faster than DuckDB for 55\% of the runs under systematic drift and for no run on JCC-H and on the attack databases. The initial GenDB-V code accepts every valid database, but it is faster than DuckDB for only 10\% of the runs on the attack databases and for 71\% to 84\% of the runs on the held-out databases. The robust code is faster than DuckDB for 93\% to 100\% of the runs. Without a robustness step, the specialized code of both methods loses its advantage over DuckDB in many runs when the data changes.

\noindent\textbf{Correctness beyond tests.}
Passing tests does not prevent silent wrong answers on other valid inputs.
We inspect Bespoke OLAP's generated code and construct databases that satisfy the declared types and keys but violate assumptions embedded in the code.
Execution confirms wrong answers for 10 TPC-H templates and all 6 SEC-EDGAR templates.
On TPC-H, valid line prices overflow 32-bit arithmetic (Q1, Q6, Q14, Q15), and an order with 70 lines exceeds a limit of 64 lines (Q21).
The code also assumes that shipment dates follow order dates (Q7), that nation names are unique (Q7, Q9, Q11), that nation keys are below 32 (Q2), and that every order has nonzero revenue (Q10).
On SEC-EDGAR, negative fiscal years and industry codes, which the declared types permit, produce wrong answers (Q2, Q3, Q4, Q6).
The code mishandles custom tags defined in another filing (Q4, Q24), miscounts a group with more than $2^{24}$ rows (Q1), and represents an empty result by a large negative constant that valid values can fall below (Q2, Q6).
Two failures need no constructed database.
On the original SEC-EDGAR database, an unseen currency binding of Q6 produces a group sum below that constant, and the code returns 21 of 22 rows.
On the held-out databases, the code returns five wrong results for Q11 with a threshold below the range of \texttt{qgen}, for example 308,064 rows where DuckDB and Umbra agree on 273,779 rows.
GenDB-V's proofs cover every input within the declared domain.
Its implementations agree with DuckDB on all tested databases.
% that expose wrong results in Bespoke OLAP.
% Testing checks selected inputs; formal verification establishes correctness throughout the declared domain.

\subsection{Optimization Case Studies}
\label{sec:eval-cases}

% We present one case for each GenDB-V setting.

\noindent\textbf{Replace self-joins with per-order checks (robust code, Q21).}
GenDB-V evaluates Q21's correlated existence tests within each order.
This avoids the two joins that the inspected DuckDB and Umbra plans use: a semi-join for \texttt{EXISTS} and an anti-join for \texttt{NOT EXISTS}.
One main scan over rows grouped by order uses bit operations to identify late deliveries and reject orders that cannot qualify, and only candidate orders require the remaining checks.
The case also shows what robust code synthesis changes.
The initial code requires that \texttt{lineitem} is sorted by order key and otherwise falls back to a serial nested-loop self-join, so many of its runs on the attack databases do not finish within 60\,s.
The robust code requires only that the rows of each order are adjacent, and its fallback is a parallel hash-based algorithm with linear cost.
The proofs establish that every path satisfies the same SQL semantics.
On TPC-H SF100, the robust code takes 284\,ms at 32 threads, outperforming DuckDB by $11.8\times$ and Umbra by $3.8\times$.

\noindent\textbf{Search in result order (joint optimization, Q2).}
For Q2, the generated storage holds supplier row positions in descending account-balance order.
The query follows that order, checks the region and part predicates, and verifies the minimum-cost condition for qualifying candidates.
Once at least 100 candidates qualify and the account balance decreases, execution stops.
The proof establishes that no unvisited candidate can enter the result, including under the query's tie-breaking rules.
The implementation takes 0.73\,ms, compared with 165\,ms for the initial plain-array implementation.

\noindent\textbf{Count matches without constructing join results (general engine, Q13).}
The engine combines a left join and aggregation into direct counting.
For Q13, the engine counts qualifying orders in each customer's index range, avoiding construction of customer--order pairs and subsequent hash grouping.
The proof preserves the distinction between \texttt{COUNT(*)} and \texttt{COUNT(o\_orderkey)}: a customer with no matching orders contributes one outer-join row but zero non-null order keys.
The engine selects this optimization from the input query's structure and checks its prerequisites at runtime.
The optimization therefore applies beyond Q13, with a general execution path for queries that do not meet the prerequisites.

\subsection{Synthesis Cost and Artifact Size}
\label{sec:eval-cost}

% Engine cost: run 2 to v016 $1,228.59 + continual evolution $363.67 + SEC-EDGAR stage $285.71 + one-engine run $954.69 = $2,832.66 (EN1/ENGINE-RESULTS.md section 3).
% Plain-array TPC-H templates, agent-written code (src/agent), same classifier (EV1/template_lines.py, 2026-11-29): median 4,743 lines initial, 9,198 robust (median growth 1.94x); specification, contract and proof 76.0% initial, 70.1% robust.
% Engine size: v020 agent/ code, 122 files, 153,443 counted lines, 67.2% specification, contract and proof (EV1/engine_v020_lines.py with tools/artifact-analysis/verus_lines.py; the same script reproduces v016: 103,955 lines, 67.5%).

% GenDB-V automates the development of verified query implementations and engines.
Synthesis for TPC-H costs \$1,285 over plain arrays, a further \$1,934 for robust code synthesis (a median of \$76 per template), and \$1,868 with joint optimization.
For SEC-EDGAR, it costs \$390 over plain arrays and \$442 with joint optimization.
Joint optimization takes a median of 5.4 and 3.5 hours per template on the two workloads.
The general engine costs \$2,833.
Specifications, contracts, and proofs account for most of the generated code.
They comprise 70\% of a robust plain-array TPC-H implementation (median 9,200 lines, about twice the initial implementation), 77\% of a jointly optimized implementation (median 6,500 lines for TPC-H and 5,900 for SEC-EDGAR), and 67\% of the general engine (153,000 lines).

% Line counts exclude comments and blank lines.

% Prior verified systems such as seL4 required approximately 20 person-years of effort~\cite{klein2009sel4}.
% Although the systems differ in scope, GenDB-V demonstrates a promising step toward reducing the human effort required to develop formally verified software through automated code and proof synthesis.

%% file: figures/eval-robustness-table.tex
% Given database: GenDB-codex .lane-tmp/general-engine/RB1/given.json (198 runs on TPC-H SF100; 2026-11-29):
% DuckDB 1,309.8 s, 47 very slow, worst 29.8x; Umbra 566.7 s, 9, 15.6x; initial 165.1 s, 0, 4.16x; robust 181.7 s, 0, 4.26x.
% All 36 cells are computed by one script from the published files results/tpch-robust/{given-tpch-sf100,attack,held-out}/times.json (GenDB-codex .lane-tmp/general-engine/RQ1/root/table2.py, 2026-12-01), fastest of the four printed systems per run. Attack: 798.6 s / 45 / 19,081.6; 930.5 / 21 / 254.2; 31,250.0 / 595 / 21,371.3; 212.6 / 0 / 305.9. Held-out: 4,018.1 / 26 / 1,628.7; 3,442.8 / 22 / 10,702.2; 13,827.2 / 1,215 / 2,083.8; 141.474 / 0 / 48.5. Corrected 2026-12-01: the attack cells of DuckDB, Umbra and the initial code now use the published robust times (second-session code for seven templates) as the fastest system; held-out robust 141.474 s prints as 141.
\begin{table}[t]
\centering\small
\caption{Total execution time in seconds (Time), number of very slow runs (Slow), and worst slowdown against the fastest system on the same run (Worst), on the given database (198 runs), the attack databases (750 runs), and the held-out databases (5,060 runs).}
\label{tab:eval-robust}
\setlength{\tabcolsep}{0.7pt}
\begin{tabular*}{\columnwidth}{@{\extracolsep{\fill}}lrrrrrrrrr@{}}
\toprule
 & \multicolumn{3}{c}{Given} & \multicolumn{3}{c}{Attack} & \multicolumn{3}{c}{Held-out} \\
\cmidrule(lr){2-4}\cmidrule(lr){5-7}\cmidrule(l){8-10}
System & Time & Slow & Worst & Time & Slow & Worst & Time & Slow & Worst \\
\midrule
DuckDB & {\footnotesize 1,310} & {\footnotesize 47} & {\footnotesize 30$\times$} & {\footnotesize 799} & {\footnotesize 45} & {\footnotesize 19,082$\times$} & {\footnotesize 4,018} & {\footnotesize 26} & {\footnotesize 1,629$\times$} \\
Umbra & {\footnotesize 567} & {\footnotesize 9} & {\footnotesize 16$\times$} & {\footnotesize 930} & {\footnotesize 21} & {\footnotesize \textbf{254$\times$}} & {\footnotesize 3,443} & {\footnotesize 22} & {\footnotesize 10,702$\times$} \\
GenDB-V initial & {\footnotesize \textbf{165}} & {\footnotesize \textbf{0}} & {\footnotesize \textbf{4$\times$}} & {\footnotesize 31,250} & {\footnotesize 595} & {\footnotesize 21,371$\times$} & {\footnotesize 13,827} & {\footnotesize 1,215} & {\footnotesize 2,084$\times$} \\
GenDB-V robust & {\footnotesize 182} & {\footnotesize \textbf{0}} & {\footnotesize \textbf{4$\times$}} & {\footnotesize \textbf{213}} & {\footnotesize \textbf{0}} & {\footnotesize 306$\times$} & {\footnotesize \textbf{141}} & {\footnotesize \textbf{0}} & {\footnotesize \textbf{49$\times$}} \\
\bottomrule
\end{tabular*}
\end{table}

%% file: figures/robustness-figure.tex
\begin{figure*}[t]
\centering
\includegraphics[width=\textwidth]{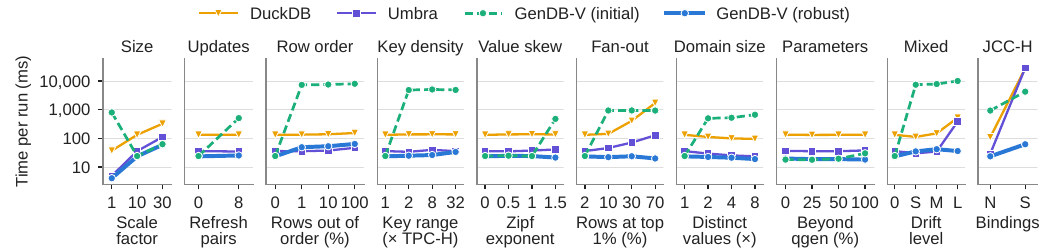}
\caption{Mean execution time per run on the held-out databases (32 threads). The first panel varies the scale factor, the Updates panel applies TPC-H refresh functions, and the JCC-H panel shows its normal (N) and skewed (S) bindings. Each other panel changes one property of TPC-H SF10, starting from the unchanged database. Fan-out is the share of child rows at the top 1\% of parents. Mixed databases change all properties except size by a small (S), medium (M), or large (L) amount.}
\label{fig:eval-robust}
\end{figure*}

%% file: figures/generalization-figure.tex
\begin{figure}[t]
\centering
\includegraphics[width=\columnwidth]{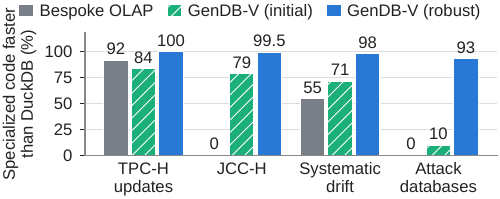}
\caption{Percentage of runs in which the specialized code is faster than DuckDB. DuckDB fallbacks do not count.}
\label{fig:eval-generalization}
\end{figure}

%% file: sections/07-related.tex
\section{Related Work}
\label{sec:related}

% \noindent\textbf{Compilation-Based Query Code Generation.}
% Query compilation produces specialized execution code through developer-defined generation and optimization rules. HIQUE~\cite{krikellas2010hique} instantiates code templates, HyPer~\cite{neumann2011compiling} generates data-centric pipelines, and LegoBase~\cite{klonatos2014legobase} compiles high-level engine implementations into specialized C code. LingoDB~\cite{jungmair2023suboperators} exposes query optimization and compilation through MLIR dialects and declarative sub-operators. Castor~\cite{feser2020castor} jointly specializes data layouts and query execution through deductive synthesis. GenDB-V lets agents develop storage designs, execution algorithms, and proofs beyond a predefined collection of implementations and transformation rules. Fixed formal synthesis targets constrain correctness while leaving implementation choices open.

\noindent\textbf{LLM-Based SQL Code Synthesis.}
CodexDB~\cite{trummer2022codexdb} and GenesisDB~\cite{trummer2025genesisdb} focus on customizable query-processing code rather than performance optimization. GenDB~\cite{lao2026gendb,lao2026gendbdemo} and Bespoke OLAP~\cite{wehrstein2026bespoke} target performance by synthesizing specialized storage and query implementations, with correctness tested against a reference system. Both only generate implementations for given SQL queries or templates. GenDB-V requires formal proofs and supports both template implementations and a general-purpose engine. Our audit and tests found incorrect results on valid inputs in Bespoke OLAP's implementations for 16 of 28 templates, illustrating the limits of testing-based validation.
% JetStream facts from arXiv:2610.20527v3 (4 Nov 2026), read 2026-11-30 (GenDB-codex .lane-tmp/general-engine/EV1/jetstream-latest): one accelerator per query template, host DBMS answers other queries (Sec. 1, 2); verified accelerators proved relative to trusted input contracts on the host (Sec. 5, Table 2); QSem "handles NULLs wherever two-valued logic agrees with SQL's three-valued logic, and refuses everything else" (Sec. 5.1); stateful accelerators keep "partial aggregates organized by the template's parameters" (Sec. 2); "Verified accelerators currently run single-threaded" (Sec. 5); TPC-H SF=20, 40 threads: stateless 1.96x over DuckDB, stateful 584x tested and 89.1x verified, GenDB 3.65x (Fig. 5). Version 1 (14 Oct) required NOT NULL columns and reported 2.1x stateless.
JetStream~\cite{balasingam2026jetstream} is concurrent work that synthesizes query-specific accelerators inside an existing DBMS, tested or formally verified. GenDB-V differs in three ways. (1) JetStream verifies one accelerator per SQL template and leaves all other queries to the host DBMS. GenDB-V also synthesizes a verified general-purpose engine for ad-hoc queries. (2) JetStream's proofs assume that the host DBMS supplies correct inputs, and its specifications handle NULL values only where two-valued logic agrees with SQL's three-valued logic. GenDB-V proves storage preservation together with query correctness, and SQLSpec models nullable data, three-valued logic, and SQL errors. (3) JetStream's large speedups come from precomputed aggregates that it maintains under writes, which GenDB-V's storage policy prohibits. Its verified accelerators currently run single-threaded, while GenDB-V also verifies parallel execution. JetStream's public artifact depends on a modified DuckDB that the artifact does not include, which prevented a direct comparison. By its reported numbers, JetStream without such state reaches a 1.96$\times$ geometric-mean speedup over DuckDB on TPC-H, compared with 5.69$\times$ for GenDB-V's robust code over plain arrays and 51.6$\times$ with joint optimization, although the experimental settings differ. 
% JetStream's own evaluation includes GenDB~\cite{lao2026gendb}, the unverified predecessor of GenDB-V.

\noindent\textbf{Formal Verification.}
Prior work formalizes SQL primarily to verify query rewrites~\cite{chu2017hottsql}, with additional work on verified query compilers~\cite{benzaken2022dbcert}. GenDB-V connects optimized storage and execution code to formal SQL semantics. More broadly, LLMs increasingly generate proofs that tools such as Lean~\cite{demoura2021lean4}, Rocq~\cite{rocqmanual}, and Verus~\cite{lattuada2023verus} check independently. This approach has achieved substantial progress in mathematics and software verification, exemplified by AlphaProof~\cite{hubert2026alphaproof} and VeruSAGE~\cite{yang2026verusage}. GenDB-V follows this approach and extends joint code and proof synthesis to high-performance SQL engines.

%% file: sections/08-conclusion.tex
\section{Conclusion}
\label{sec:conclusion}

GenDB-V enables AI agents to synthesize formally verified database engines. Fixed formal synthesis targets and independent verification let agents jointly explore storage designs, execution code, and proofs. Robust code synthesis lets agents generalize verified code beyond the given database state. Experimental results show that formal correctness can accompany substantial speedups over established engines, offering a promising path toward AI-generated and formally verified database systems.